\documentclass{aa501}
\usepackage{graphics}
\usepackage{amssymb}
\newcommand{\pad}{.\hskip-2pt$^\circ$}
\begin{document}

\title{Clumpy outer Galaxy molecular clouds and the steepening of the IMF}

\author{J. Brand
	  \inst{1}
          \and
         J.G.A. Wouterloot
          \inst{1,2,3}
	  \thanks{\emph{Present address:}
	  Joint Astronomy Centre, 660 N. A'ohoku Place, University Park,
          96720 Hilo, Hawaii, USA}
          \and
	 A.L. Rudolph
	  \inst{4} 
	  \thanks{NSF Career Fellow}
          \and
         E.J. de Geus
          \inst{5,6}
	  \thanks{\emph{Present address:}
	  NFRA, Postbus 2, 7990 AA Dwingeloo, The Netherlands}
	   }

\offprints{J. Brand (brand@ira.bo.cnr.it)}

\institute{
Istituto di Radioastronomia, CNR, Via Gobetti 101, I-40129 Bologna, Italy 
 \and
Radioastronomisches Institut, Universit\"at Bonn, Auf dem H\"ugel 71, 
D-53121 Bonn, Germany 
 \and 
I. Physikalisches Institut, Universit\"at zu K\"oln, Z\"ulpicher Stra\ss e 77,
D-50937 K\"oln, Germany 
 \and
Department of Physics, Harvey Mudd College, Claremont, CA 91711, USA 
 \and
Department of Astronomy, University of Maryland, College Park, MD 20742, USA 
 \and
Astronomy Department, Caltech, Pasadena, CA 91125, USA
}

\date{Received ;accepted }

\abstract{We report the results of high-resolution ($\sim$0.2~pc) CO(1--0) 
and CS(2--1) observations of the central regions of three star-forming 
molecular clouds in the far-outer Galaxy ($\sim$16~kpc from the Galactic 
Center): WB89~85 (Sh~2-127), WB89~380, and WB89~437.
We used the BIMA array in combination with IRAM~30-m and NRAO~12-m 
observations. The GMC's in which the regions are embedded were studied by 
means of KOSMA~3-m CO(2--1) observations (here we also observed WB89~399).
We compare the BIMA maps with optical, radio, and near-infrared observations.
Using a clumpfind routine, structures found in the CO and CS emission are
subdivided in clumps, the properties of which are analyzed and compared with 
newly derived results of previously published single-dish measurements of 
local clouds (OrionB~South and Rosette). \hfill\break\noindent
We find that the slopes of the clump mass distributions (--1.28 and --1.49,
for WB89~85 and WB89~380, respectively) are somewhat less steep than found 
for most local clouds, but similar to those of clouds which have been analyzed 
with the same clumpfind program.
\hfill\break\noindent
We investigate the clump stability by using the virial theorem, including all
possible contributions (gravity, turbulence, magnetic fields, and 
pressure due to the interclump gas). It appears that under reasonable
assumptions a combination of these forces would render most clumps stable. 
Comparing only gravity and turbulence, we find that in the far-outer Galaxy 
clouds, these forces are in equilibium (virial parameter $\alpha \approx 1$) 
for clumps down to the lowest masses found (a few M$_{\odot}$). For clumps in 
the local clouds $\alpha \approx 1$ only for clumps with masses larger than a 
few tens of M$_{\odot}$. Thus it appears that in these outer Galaxy clumps 
gravity is the dominant force down to a much lower mass than in local clouds, 
implying that gravitational collapse and star formation may occur more readily 
even in the smallest clumps. Although there are some caveats, due to the
inhomogeneity of the data used, this might explain the apparently steeper IMF 
found in the outer Galaxy.
}

\authorrunning{Brand et al.}
\titlerunning{Clumpy outer Galaxy clouds}
\maketitle

\section{Introduction \label{intro}}

Studies of molecular clouds have spanned size scales from many tens to tenths 
of parsecs. The large-scale properties of molecular clouds can be studied 
throughout the Galaxy as well as in external galaxies. For the smallest 
structures in molecular clouds, however, lack of resolving
power has limited detailed studies to molecular clouds nearest to the Sun.
Therefore, it has never been investigated whether the effects of a different 
physical environment influences the properties of those structures. 
The possibility of making high-resolution maps with mm-interferometers allows 
a study of clump properties in clouds at much larger distances from the Sun, 
and therefore in a potentially different physical and chemical environment. 
The outer part of the galactic disk is such a region, where molecular clouds 
are more sparsely distributed (Wouterloot et al.~\cite{wbbk}), the diffuse 
galactic interstellar radiation field is weaker (Cox \& Mezger~\cite{cox}, 
Bloemen~\cite{bloemen}), the metallicity is lower (Shaver et 
al.~\cite{shaver}, Fich \& Silkey~\cite{fichsilk}, Wilson \& 
Matteucci~\cite{wilson}, Rudolph et al.~\cite{rudolph97}), and the cosmic-ray 
flux density is smaller (Bloemen et al.~\cite{bloemenben}), compared to the 
solar neighbourhood. 

\noindent
Previously, we have studied molecular clouds at galactocentric distances
$R>$15~kpc (Brand \& Wouterloot~\cite{brandw1}) and analyzed molecular cloud
properties across the Galaxy (Brand \& Wouterloot~\cite{brandw2}, hereafter
BW95). It was found that cloud kinetic temperatures as well as CO luminosities
are similar to GMCs of the same mass in the solar neighbourhood.
These results were in contradiction with those of Mead \& Kutner
(\cite{mead1}), who derived kinetic temperatures of 7~K for a sample of
clouds at $R\sim$13~kpc, significantly colder than GMCs in the solar
neighbourhood, and Digel et al. (\cite{digel}), who found that outer Galaxy
molecular clouds appear to be underluminous in CO with respect to their
virial mass, by a factor of four, compared to local clouds (suggesting that
the CO-to-H$_2$ conversion factor, $X$, should be four times higher). BW95
showed that the Digel et al. result is a consequence of the small
number of clouds studied by them.
BW95 also found that outer Galaxy clouds are generally less massive than
inner Galaxy clouds, and that in the inner Galaxy there are relatively more
large clouds (see also May et al.~\cite{may}); they furthermore noted
that outer Galaxy clouds have larger radii than inner Galaxy
clouds of the same mass, which in part could be explained by a lower
pressure of the surrounding ISM at large $R$, allowing the clouds to settle 
at larger equilibrium radii.  Wouterloot et al. (\cite{wouterfieg}), and Fich 
\& Terebey (\cite{fichter}), using far-infrared luminosities of IRAS point
sources in the outer Galaxy determined by Wouterloot \& Brand (\cite{wb89};
hereafter WB89), measured a slope for the initial mass function in the outer 
Galaxy that is {\sl steeper} than the IMF measured in the solar neighbourhood. 
A similar indication of steepening has been found by Garmany et al. 
(\cite{garmany}) from O-stars (M$>$20~M$_{\odot}$) within 2.5~kpc from the 
Sun. These differences are generally attributed to the different conditions 
in the outer Galaxy. 
On the other hand Casassus et al. (\cite{casassus}), from a study of IRAS 
sources with the colours of UCH{\sc ii} regions concluded that the exponent of 
the IMF does not seem to vary.

\smallskip\noindent
The aim of the present study is to investigate the properties of molecular
clumps (sizes from $0.2-2$~pc) in GMCs in the outer Galaxy, and to explain
any differences with clump properties in local clouds in terms of effects due
to a different physical environment. At the typical distances for these
objects of 10~kpc, the resolution required for such a study is $\sim$ 5\arcsec,
which can only be obtained by using an interferometer. We employed the BIMA
mm-interferometer (Welch et al.~\cite{welch}) to map CO(1--0) and CS(2--1) in 
three molecular clouds at $R>$15~kpc. To compensate for the lack of
zero-spacing information, which is particularly important for structures that
are extended over size-scales larger than 1\arcmin, the data were 
complemented with observations from the Kitt Peak 12-m (CO) and the IRAM 30-m
telescopes (CS). The combined maps allow us to compare clump properties in the 
far outer Galaxy with those obtained from single dish observations of nearby 
GMCs.
Three clouds (2 of which were also observed with BIMA) were mapped with
KOSMA, to allow derivation of their masses. 

Sect.~\ref{obs} describes the sample of objects, the observations and the
data reduction.  In Sect.~\ref{results} we present the results for
the individual sources, while in Sect.~\ref{clumpana} we describe the method
used to identify clumps from the data. Sect.~\ref{discuss} is devoted to a
discussion of the clump physical properties and a comparison with local
clouds. In particular, in Sect.~\ref{clumpstab} we look at the implications
of our observational results on the slope of the IMF as a function of $R$.
In Sect.~\ref{summ} we summarize the main results.

\begin{table*}
\caption[]{BIMA observational parameters.
\label{obspar}}
\begin{flushleft}
\begin{tabular}{rcccrrrrrr}
\hline\noalign{\smallskip}
\multicolumn{1}{c}{Source}& \multicolumn{1}{c}{IRAS}&
\multicolumn{1}{c}{$\alpha$(1950)}&  \multicolumn{1}{c}{$\delta$(1950)}&
\multicolumn{1}{c}{Molecule}& 
\multicolumn{1}{c}{Array} & beam & pixel & rms & K/Jy \\ 
WB89 & & {\sl h\ m\ s} & \degr\ \arcmin\ \arcsec & & & \arcsec & \arcsec 
& Jy & \\
\hline\noalign{\smallskip}
85  &21270+5423& 21 27 05.9 & +54 23 42 & CO(1--0) & B,C & 10.7$\times$8.5   
& 2   & 1.2 & 1.02 \\
380 &01045+6505& 01 04 35.7 & +65 05 21 & CO(1--0) & B,C & 7.2$\times$4.7    
& 1.5 & 1.2 & 2.73\\
    &          &            &           & CS(2--1) & B,C & 8.9$\times$6.6    
& 2   & 0.46 & 2.19\\
437 &02395+6244& 02 39 30.6 & +62 44 22 & CO(1--0) & C   & 21.8$\times$14.9  
& 3   & 1.9 & 0.28\\
    &          &            &           & CS(2--1) & C   & 19.3$\times$16.0  
& 3   & 0.65 & 0.41 \\
\noalign{\smallskip}
\hline
\end{tabular}
\end{flushleft}
\end{table*}

\section{Observations and Data Reduction \label{obs}} 

\subsection{Source Selection}

The WB89-catalogue presents an extensive CO survey towards IRAS point sources
located in the outer Galaxy (second and third quadrants), with infrared colours
that discriminate in favour of sources frequently associated with H$_2$O masers
and dense molecular cloud cores (Wouterloot \& Walmsley~\cite{wouterwalms}),
and hence with star forming regions.  We selected the four sources at 
$R >$15~kpc with the highest far-IR luminosity, to wit \object{WB89~85} 
(\object{IRAS~21270+5423}), \object{WB89~380} (\object{IRAS~01045+6505)}, 
\object{WB89~399} (\object{IRAS~01420+6401}), 
and \object{WB89~437} (\object{IRAS~02395+6244}). Unless specified otherwise, 
in the following we shall for convenience use the WB89-name to indicate the 
molecular clouds associated with these IRAS sources, rather than the IRAS
source itself.

\subsection{Single-Dish Observations} 

{\sl NRAO 12-m}

\noindent
WB89~85, 380, and 437 were observed using position switching on a 9$\times$9 
grid with 30\arcsec\ spacing in CO J=1--0 with the Kitt Peak 12-m
telescope on December 4-9, 1991. The emission was found to extend beyond this 
region in all three cases; subsequently the GMC near WB89~85 was completely 
mapped. The telescope beamwidth at 115~GHz is 56\arcsec. The standard chopper 
wheel method (Ulich \& Haas~\cite{ulich}, Kutner \& Ulich~\cite{kutnerul}) was 
used for calibration. All intensities are on a a $T_{\rm R}^*$-scale. 
System temperatures were typically 350 -- 500~K.
We used filterbanks of 128 channels of 100~kHz for each of the polarizations. 
The velocity coverage is therefore 33~kms$^{-1}$ with 0.26~kms$^{-1}$ 
resolution. After averaging, the rms noise level is $0.13-0.25$~K. 

\noindent
We have also observed $^{12}$CO(2--1), $^{13}$CO(1--0), CS(2--1), 
HCO$^+$(1--0), and HCN(1--0) at various locations on and near the three main
CO peaks surrounding IRAS21270+5423.

\smallskip\noindent
{\sl IRAM 30-m}

\noindent
WB89~85, WB89~380, and WB89~437 were observed in CS J=2--1, 3--2, and 5--4
with the IRAM 30-m telescope at Pico Veleta from July 24 -- 26 1991. The maps 
are on a 11$\times$11 grid with 15\arcsec\ spacing, covering all CS emission. 
The telescope beamwidth at these frequencies (98, 147, and 245~GHz) is 
respectively 26\arcsec, 16\arcsec, and 11\arcsec. The standard chopper wheel 
method was employed for temperature calibration. All intensities are on a 
$T_{\rm A}^*$-scale. System temperatures were typically
200 -- 250~K. Using a 128 channel filterbank of 100~kHz per channel,
the velocity coverage is 40~kms$^{-1}$ and the resolution 0.3~kms$^{-1}$.  
Observations were made by position switching; the typical rms noise level in 
the spectra is 0.08~K. Low-order polynomial baselines were subtracted. 
In addition, 15 more outer Galaxy clouds were observed in the same 
transitions, to search for potentially interesting objects. With a few 
exceptions, these were all single-pointed observations at the IRAS PSC 
position. 

\smallskip\noindent
{\sl KOSMA 3-m}

\noindent
WB89~380, WB89~399, and WB89~437 were observed in $^{12}$CO(2--1) in
August 1998 and September 1999 with the refurbished KOSMA 3-m telescope (see
Kramer et al.~\cite{kramertel}). The beamsize at 230~GHz is about 1\farcm9 and
the clouds were observed in position switching mode
on a 1\arcmin\ raster in galactic coordinates, covering the region where
Brand \& Wouterloot (\cite{brandw1}) detected $^{12}$CO(1--0) at the lower
resolution (and raster) of about 4\arcmin.
We used an acousto-optical spectrometer with a channel spacing of 167~kHz
and effective resolution of 360~kHz (0.47~kms$^{-1}$). The sky transmission
was estimated by measuring the radiation temperature of the blank sky at the
elevation of the sources. Analogous to the standard chopper wheel calibration,
the intensities were corrected to the $T_{\rm A}^*$-scale. The rms noise level
in the spectra is about 0.10 -- 0.15~K. For the mass calculations (see
Sect.~\ref{results}) we used $T_{\rm mb}$ ($\eta_{\rm mb} = 0.7$).

\subsection{Interferometer Observations}

WB89~85, 380, and 437 were observed with the three-element BIMA
interferometer in 1991 October, May and June, respectively. CO(1--0) 
observations of WB89~85 and WB89~380 were made with the B- and C-array, and of 
WB89~437 with the C-array. CS(2--1) observations of WB89~380 were made with 
the B- and C-array, and of WB89~437 with the C-array.
Flux- and phase calibrators were 3C84, BL~Lac, and 3C454.
The primary beam of the array is 100\arcsec\ at 115~GHz and 117\arcsec\ at
98~GHz. Table~\ref{obspar} lists the parameters of the observations.
The instrumental phase and gain were determined with the standard BIMA data
reduction package Miriad.

\begin{figure}
\resizebox{\hsize}{!}{\includegraphics{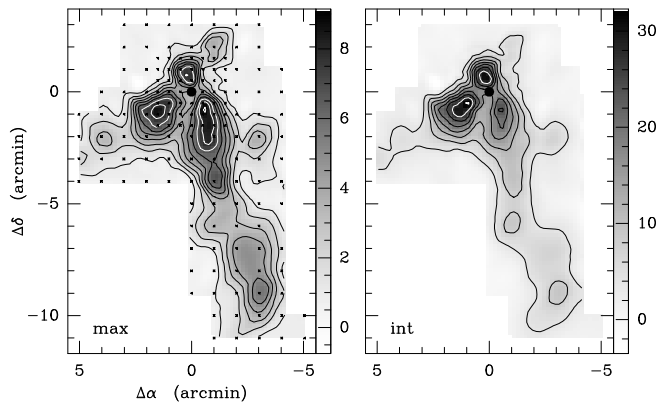}}
 \caption{NRAO 12-m map of CO peak $T_{\rm R}^*$ (left) and
$\int T_{\rm R}^*{\rm d}v$ (right) for 
$-100 < V_{\rm lsr} < -88 $~km$\,$s$^{-1}$ for WB89~85. Contour values are
1(1)9~K (left) and 4(4)32~Kkm$\,$s$^{-1}$ (right). 
The IRAS source position is indicated with the filled circle at (0,0), while
the small crosses indicate the observed positions.}
\label{fig1}
\end{figure}

\subsection{Combining Single-Dish and Interferometer Data}

In all three sources the distributions of CO and CS are extended with respect
to the BIMA synthesized beams, which implies that the interferometer
observations are not sensitive to a significant fraction of the emission, due
to a lack of information on the low spatial frequencies. This information has
been obtained using the single-dish observations.  The maps from the 12-m
sample spatial frequencies from 0 to 4.5 k$\lambda$ in the {\it uv}-plane, and
those from the 30-m sample spatial frequencies from 0 to 9.6 k$\lambda$.

\noindent
The method used for this data combination is similar to that described by 
Vogel et al. (\cite{vogel}) and Bieging et al. (\cite{bieging}). 
A single dish observation is a convolution of the true spatial
distribution of the emission with the beam. To take out the effect of
this convolution, the maps were deconvolved using the CLEAN-algorithm.
Similarly, an interferometer observation is a convolution of the true
distribution of the emission with the primary beam of each of the BIMA
telescopes. The deconvolved single-dish maps were convolved with the
primary-beam pattern of the BIMA telescopes. To ensure correct
scaling and positioning of the single-dish data with respect to the
interferometer observations, maps were constructed from the interferometer 
and single dish {\it uv}-data separately covering {\sl only} the overlapping 
part in {\it uv}-space. For the 12-m and BIMA comparison this involved a 
{\it uv}-range of 2.3 to 4.0~k$\lambda$, and for the 30-m and BIMA comparison 
we used 2.0 to 7.0 k$\lambda$. In these maps we found small offsets in the 
absolute positions between the single dish and the interferometer map centers. 
These were 10\arcsec\ or less for the comparison with the 12-m maps, and 
6\arcsec\ or less for the 30-m maps. Since the pointing accuracy of the 
single-dish telescopes is worse than that of the interferometer, we corrected 
the central coordinates of the single dish maps to match those of the 
interferometer observations. The shifted maps made from the same limited 
{\it uv}-range were then used to derive the relative scaling of 35~Jy/K for
the 12-m CO observations, and 9~Jy/K for the 30-m CS(2--1) observations. These
values are within 5\% of the theoretical gain of the single-dish antennas. 

\begin{figure*}
\resizebox{12cm}{!}{\includegraphics{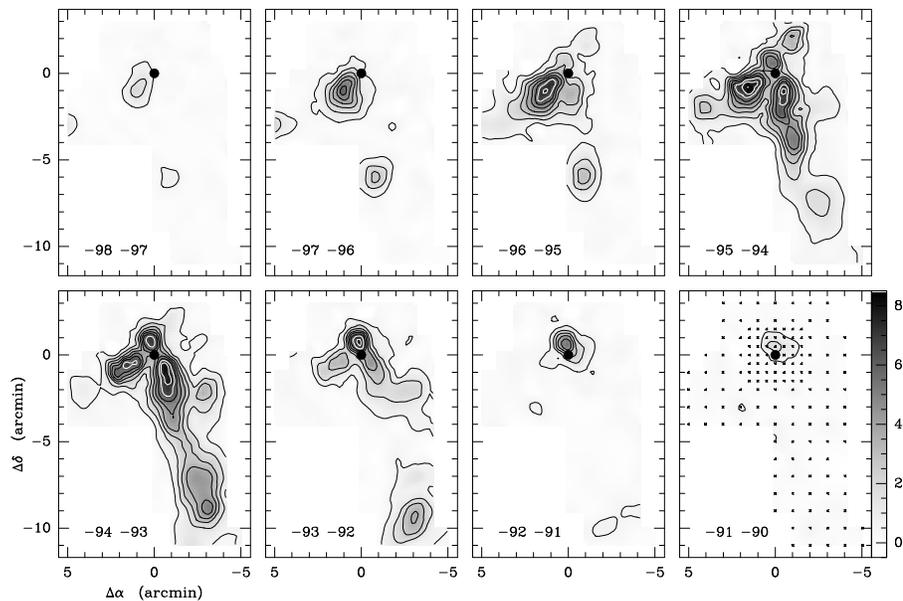}}
 \hfill
  \parbox[b]{55mm}{
 \caption{NRAO 12-m $^{12}$CO(1--0) map of the WB89~85 cloud, showing
channel maps in 1~kms$^{-1}$ intervals. Contour values are
0.5(1)8.5~Kkms$^{-1}$. Symbols as in Fig.~\ref{fig1}}
\label{wb85kpchan}
}
\end{figure*}

\begin{figure*}
  \resizebox{12cm}{!}{\includegraphics{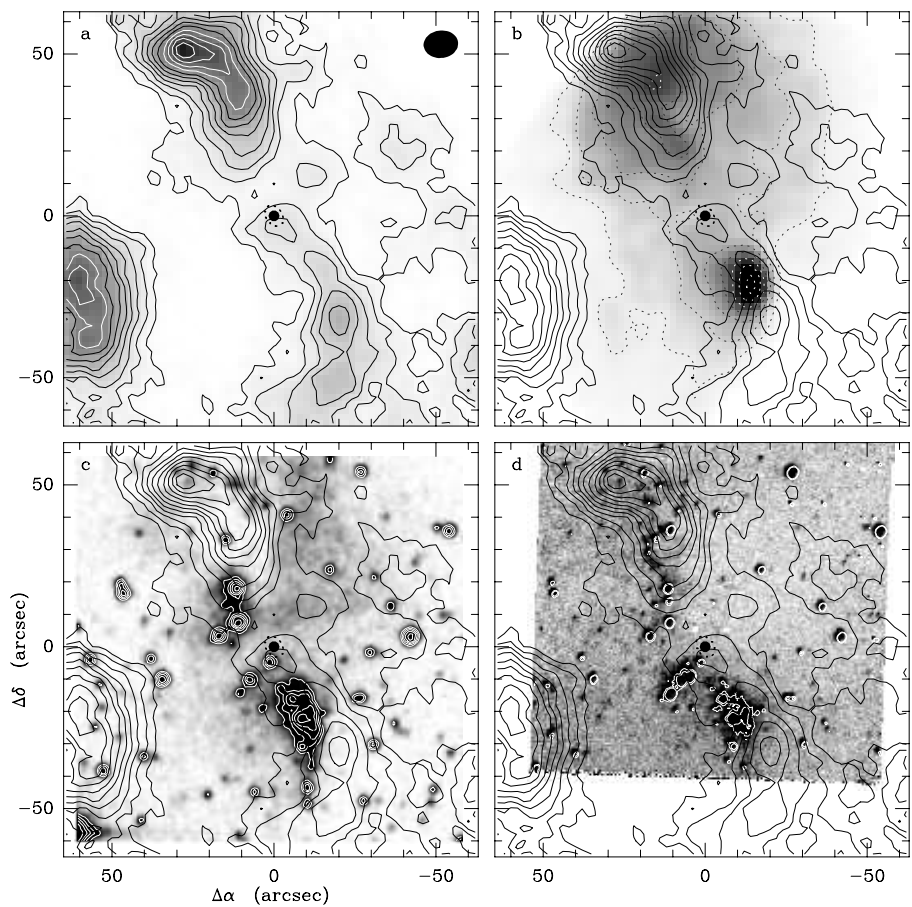}}
 \hfill
  \parbox[b]{55mm}{
  \caption{{\bf a}\ High-resolution BIMA CO image of the WB89~85 region.
The emission is integrated over $-98.25 < V_{\rm lsr} <
-89.50$~km$\,$s$^{-1}$, and the (primary beam-corrected) 
contour levels are 14(14)~Kkms$^{-1}$. The resolution of the BIMA synthesized 
beam is indicated in the upper righthand corner. The filled circle is the IRAS
source, the thick dotted contour is the associated error ellipse.
{\bf b}\ The BIMA CO contours superimposed on the distribution of the 6-cm 
continuum radiation (halftone and dotted contours), with a peak flux in the 
map of about 170~mJy/beam. 
{\bf c}\ As {\bf b}, but with the BIMA CO contours superimposed
on the optical image from the DSS. For clarity, the more intense optical
emission has been contoured (white). {\bf d} As {\bf c}, but with the BIMA CO 
contours superimposed on the K-frame obtained by de Geus \& E. Lada
(pers. comm.). The cuts of the gray scale have been set to show the
low-level emission; for clarity, the more intense K-emission has been 
contoured (white; the lower levels only, to avoid confusion).}
\label{s1276cmco}
}
\end{figure*}

\noindent
After applying the shift and gain, the single-dish maps were (fast)
Fourier-transformed to {\it uv}-space and sampled on a set of eight circular 
tracks in the {\it uv}-plane from 0 to 4.0~k$\lambda$ for the 12-m maps and 
from 0 to 8.0~k$\lambda$ for the 30-m maps. To check this procedure, the 
{\it uv}-datasets acquired in this way were Fourier transformed back into map 
space; the resulting maps as well as the beam profiles were compared to the 
input maps and beam patterns. The maxima and the total fluxes of the maps 
agree to with 3\%, and Gaussian fits to the beam maps give beam profiles that 
are within 2\% of the nominal sizes of the single-dish beams. 

\noindent
The single-dish {\it uv}-data were then combined with the {\it uv}-datasets 
from the interferometer. Complete data cubes were subsequently obtained by 
Fourier transforming the combined {\it uv}-datasets to the map plane, using 
the Miriad routine INVERT, with natural weighting (which is essential to
maintain the appropriate relative weights of the single-dish and interferometer
contributions). From the resulting cubes the ``dirty'' beam was deconvolved
using the Miriad routine CLEAN, and convolved with a ``clean''-beam
using RESTORE. From the separate planes in the data cubes the CLEAN and RESTORE
procedure left only the inner quarter with useful data, which was extracted
from the cubes. In order to correct for the response of the primary beam of the
interferometer elements, these data cubes were divided by the normalized
primary beam profile. This final procedure allows us to determine clump
properties confidently over the whole image, but has the important disadvantage
that the noise is not uniform over the image. This leads to uncertainties in
the clump finding process, which will reflect on the completeness of the clump
search, particularly for the weakest (i.e. smallest and least massive) clumps.

\begin{figure*}
  \resizebox{12cm}{!}{\includegraphics{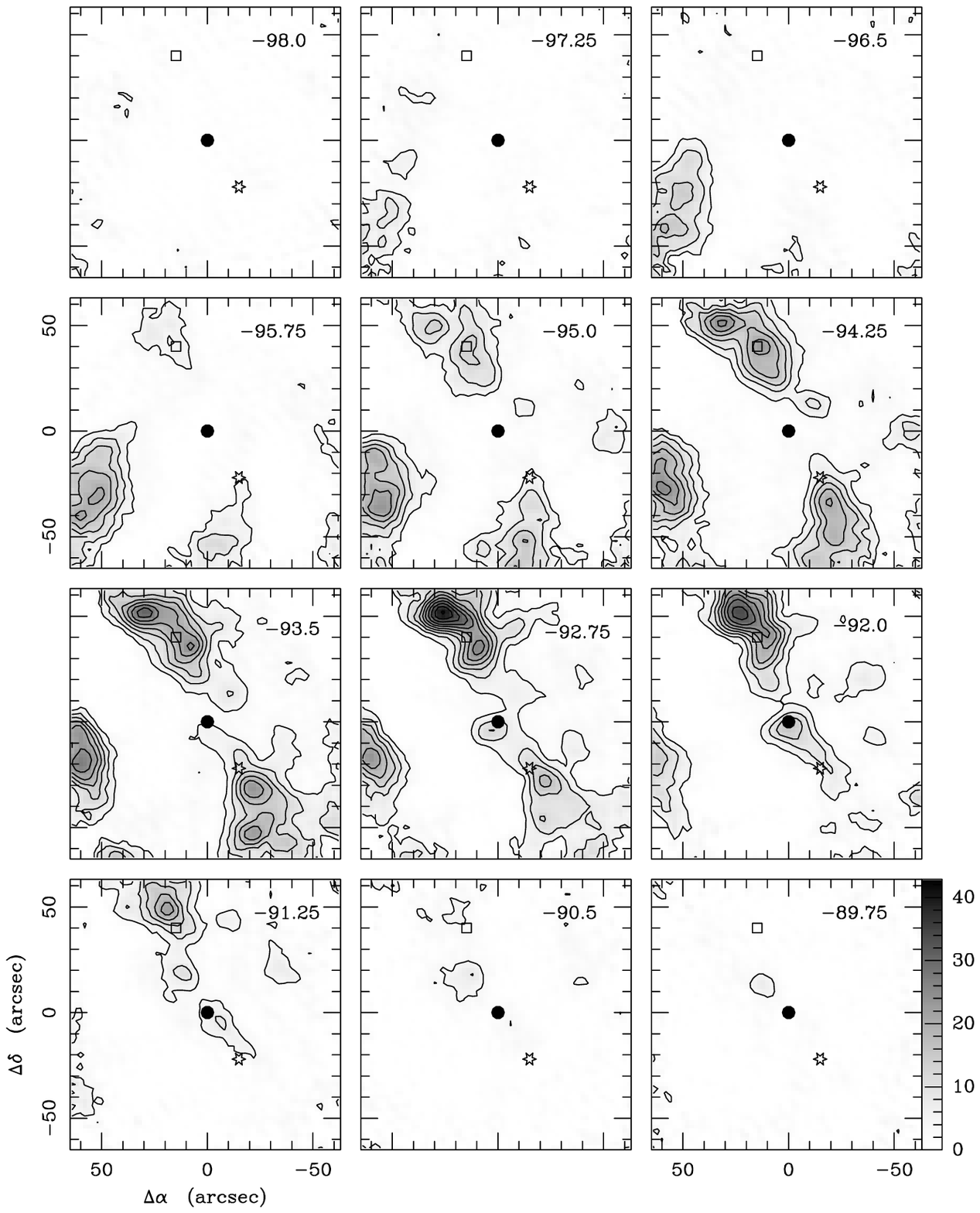}}
  \hfill
  \parbox[b]{55mm}{
  \caption{High-resolution $^{12}$CO(1--0) channel maps of the Sh~2-127 region
showing the velocity structure between --98.0 and --89.75~km$\,$s$^{-1}$.
Each panel shows the emission in a 0.75~kms$^{-1}$
wide interval, centered at the indicated velocity. The position of the IRAS
source WB89~85 is indicated by the filled circle and that of the compact
H{\sc {ii}} region/radio component B by the asterisk. The square denotes the
position of the peak of radio component A (see text). Contour levels are
4(4)~Kkms$^{-1}$.}
\label{fig2_chan}}
\end{figure*}

\section{Results \label{results}}

{\bf WB89~85} (IRAS 21270+5423) is located in \object{Sh~2-127}, which is an
optical H{\sc ii} region at a kinematic distance of 15.0~kpc from the galactic
center and 11.5~kpc from the Sun. The associated molecular cloud
was completely mapped in CO with the NRAO 12-m at Kitt Peak. Fig.~\ref{fig1}
shows the NRAO 12-m CO map of peak temperature (left) and integrated
emission (right). From the total CO luminosity, and using a value for the
$X$-factor ($N{\rm (H_2)}/W_{\rm CO}$) of 1.9$\times$10$^{20}$ (Strong \&
Mattox~\cite{strongmat}), we derive a mass $M_{\rm CO}$ (see Sect.~4) for this
molecular cloud of 1.6$\times 10^4$~M$_\odot$. The IRAS far infrared
luminosity of the point source is $L_{\rm fir}$=1.0$\times 10^5$~L$_\odot$, 
which indicates the presence of the equivalent of an O7~V$_0$ star (or 
O8-8.5~V; Panagia~\cite{panagia}). The relevant parameters of this cloud, and 
the other clouds, are collected in Table~\ref{cloudpar}. Fig.~\ref{fig1} shows
that the IRAS source is located in a relative minimum, and is surrounded by
three main clouds. From the channel maps in Fig.~\ref{wb85kpchan} we see
that most of the emission is between $-$95 and $-$93~kms$^{-1}$, and that
the three clouds near the IRAS source all have the peak of their emission at 
different velocities.

\begin{table}
\caption[]{Cloud parameters.
\label{cloudpar}}
\begin{flushleft}
\begin{tabular}{rrrccc}
\hline\noalign{\smallskip}
\multicolumn{1}{c}{Source}& \multicolumn{1}{c}{$d$}&
\multicolumn{1}{c}{$R$} & \multicolumn{1}{c}{$L_{\rm fir}$} &
\multicolumn{1}{c}{$M_{\rm CO}$} & 
\multicolumn{1}{c}{$r_{\rm eff}^{\clubsuit}$} \\
WB89 &(kpc)  & (kpc) & (10$^4$~L$_\odot$) & (10$^4$~M$_\odot$) & (pc) \\
\hline\noalign{\smallskip}
 85 & 11.5 & 15.0 & 10 & 1.6$^1$ & 11.4$^1$ \\
380 & 10.3 & 16.6 & 10 & 3.5$^2$ & 22.4$^2$ \\
399 & 9.9  & 16.6 & 4.3& 2.6$^2$ & 21.3$^2$ \\
437 & 9.1  & 16.2 & 7.1& 0.94$^2$& 13.3$^2$ \\
\noalign{\smallskip}
\hline
\multicolumn{6}{l}{$^{\clubsuit}$\ Effective radius, $r_{\rm eff} = 
\sqrt{Area/\pi}$, corrected for beam size }\\
\multicolumn{6}{l}{$^1$\ From NRAO~12-m CO(1--0) data } \\
\multicolumn{6}{l}{$^2$\ From KOSMA~3-m CO(2--1) data } \\
\end{tabular}
\end{flushleft}
\end{table}

\noindent
Radio-continuum observations (Rudolph et al.~\cite{rudolph}), revealed the
presence of an extended source (their component A; deconvolved size 30\arcsec),
and a compact H{\sc ii} region (their component B; deconvolved size
9\arcsec), about 1\arcmin\ to the southwest.
Both components are connected and enveloped by low-level diffuse emission
(see Fig.~1 in Rudolph et al.), and indicate the presence of the equivalent 
of one O7 and one O8.5~V$_0$ star, for components A and B, respectively, 
consistent with the IRAS luminosity.

\noindent
Fig.~\ref{s1276cmco}a shows the distribution of the integrated CO intensity
towards Sh 2-127 obtained from the BIMA and the NRAO 12-m observations, 
corrected for the primary beam of the BIMA antennas. 
It shows the three main clouds from Fig.~\ref{fig1}, as well as some 
low-intensity emission.  

\noindent
The distribution in velocity of the CO emission is shown in
Fig.~\ref{fig2_chan}. The emission is quite structured, and several emission
peaks can be seen in each panel; in Sect.~\ref{clumpana} we use a clump
finding program that leads to the identification of 62 clumps in these data.

\begin{figure}
  \resizebox{\hsize}{!}{\includegraphics{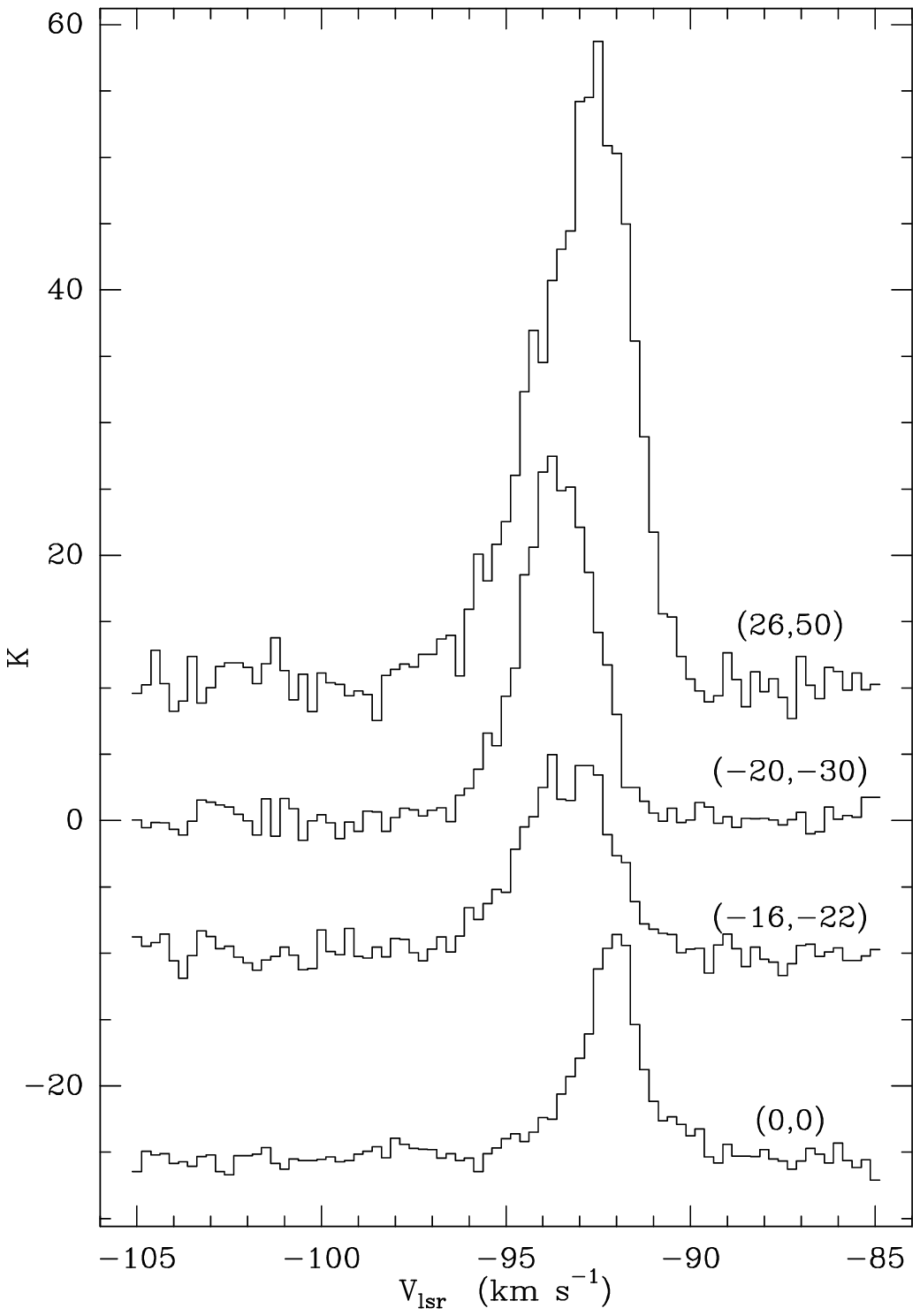}}
\caption{Spectra at four positions towards WB89~85 extracted from 
the BIMA CO datacube, averaged over 5$\times$5 pixels of 2\arcsec\ centered 
at the indicated offset from the map center. }
\label{spectra_s127}
\end{figure}

\noindent
The CO peaks in the interferometer map were searched for CS emission with the
IRAM 30-m telescope. A CS core of size (at 1--2$\sigma$ level) $\approx$
50\arcsec$\times$60\arcsec\ (2.8~pc $\times$ 3.4~pc) was detected towards the
northern CO peak, centered at offset $\sim$(17\arcsec,49\arcsec).
This corresponds to within 5\arcsec\ with both the position of the northern
radio continuum peak and the northern CO-peak.
A search with the NRAO 12-m for the higher-density tracing transitions
CS(2--1) and HCN(1--0) towards the easternmost CO peak in the single-dish CO
map has resulted in the detection of both molecules ($T_{\rm R}^* \sim 0.3$~K),
which implies densities of at least a few times 10$^4$~cm$^{-3}$ (e.g.
Evans~\cite{evans}); likewise, HCO$^+$(1--0) was searched for and detected
($T_{\rm R}^* \sim 0.5$~K) towards all three CO peaks that surround the IRAS 
position. A fit to the hyperfine lines in the HCN(1--0) sum-spectrum indicates 
that the emission is optically thin ($\tau \sim 0.6\pm 0.7$).

\noindent
In Fig.~\ref{s1276cmco}b, c, and d we compare the integrated CO distribution
with respectively the VLA 6-cm radio continuum map (Rudolph et
al.~\cite{rudolph}), the optical image, taken from the Digital Sky Survey 
(DSS), and a 2~$\mu$m (K-band) image, obtained with the 4-m telescope at Kitt 
Peak, by de Geus \& E. Lada (pers. comm.).

\begin{figure*}
  \resizebox{\hsize}{!}{\includegraphics{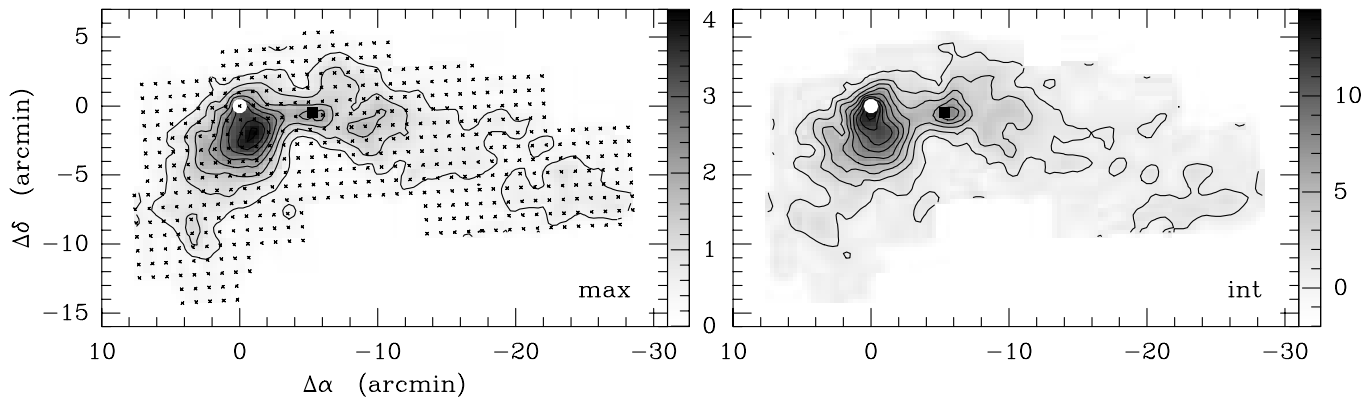}}
  \hfill
\caption[ - ]{KOSMA $^{12}$CO(2--1) map of the cloud associated with WB89~380, 
showing the peak (left)
and integrated CO intensity (right) from 
$-92 < V_{\rm lsr} < -82$~km$\,$s$^{-1}$. The lowest contour
level and contour step are 0.5 and 0.5K (left) and
1.0 and 1.5~K~km$\,$s$^{-1}$ (right). Indicated are the 
locations of two IRAS point sources that were included in Wouterloot
\& Brand (\cite{wb89}): WB89~379 (IRAS 01037+6504; square) and WB89~380 
(circle; at (0,0)). The small crosses indicate the
observed positions. The interferometer observations 
are centered on the position of WB89~380.}
\label{kos380co}
\end{figure*}

\noindent
The diffuse radio continuum emission covers much of
the area in Fig.~\ref{s1276cmco}b, and correlates (together with the 
components A and B) fairly well with that of the high-resolution CO 
emission, although there are significant differences (see below).

\noindent
The optical- and the CO emission are clearly anti-correlated. Comparison of 
Fig.~\ref{s1276cmco}b and c shows that the optical emission of component A is 
for the most part obscured by, and therefore located behind, the large 
northern CO complex: only some relatively faint emission of this component is 
optically visible, beyond the western edge of the obscuring molecular cloud, 
which shows that component A is not (completely) embedded. 
Its peak is slightly offset to the SW from the main CO peak, but coincides 
with a clump (nr. 10 in Table~\ref{s127cocl}).

\begin{figure}
  \resizebox{\hsize}{!}{\includegraphics{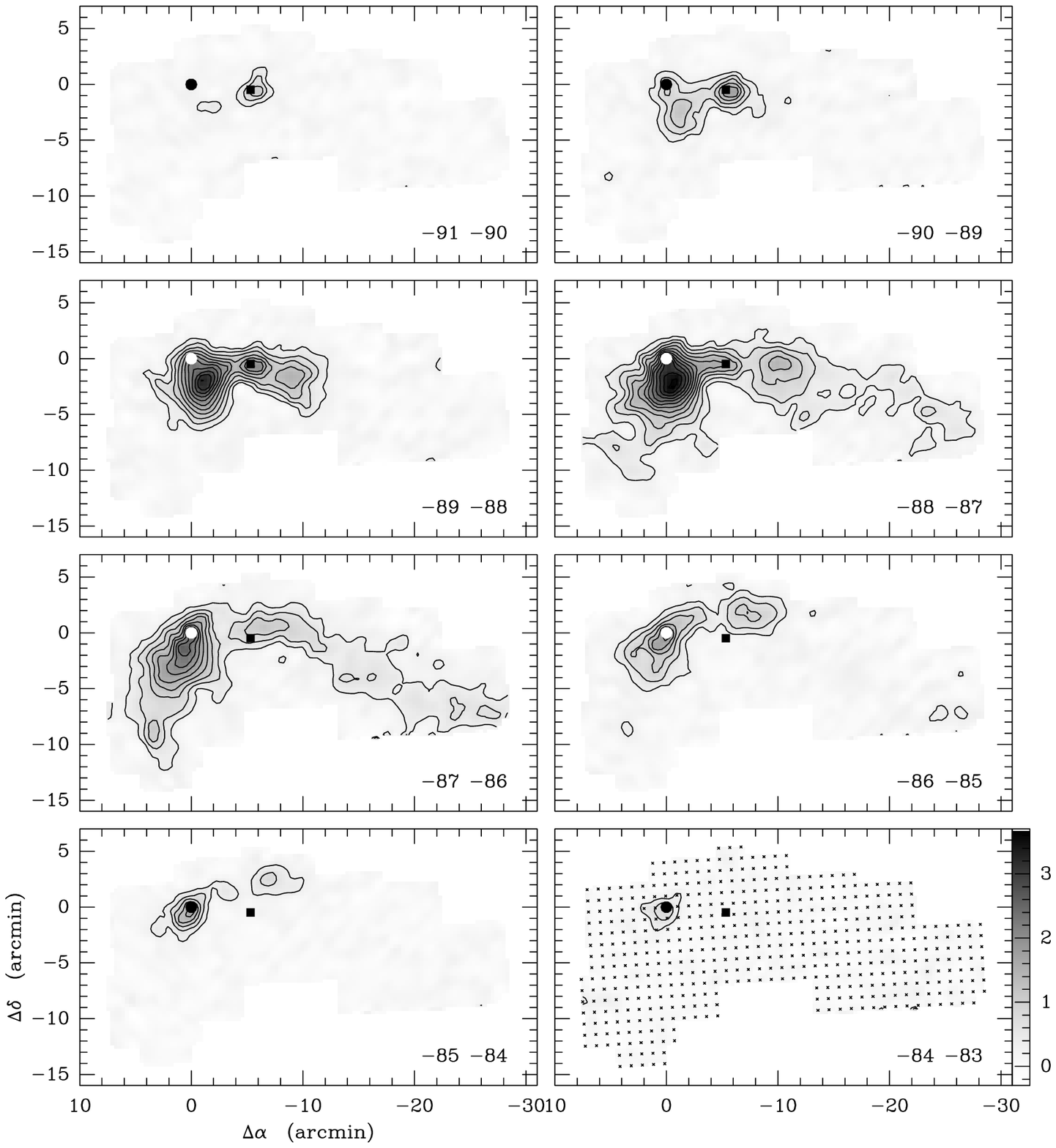}}
  \hfill
\caption[ - ]{KOSMA $^{12}$CO(2--1) map of the WB89~380 cloud, showing channel 
maps in intervals of 1~km$\,$s$^{-1}$. The lowest contour
level and contour step are 0.3~K~km$\,$s$^{-1}$. Symbols are as in 
Fig.~\ref{kos380co}.}
\label{kos380cochan}
\end{figure}

\noindent
The optically visible compact H{\sc ii} region (component B) lies at the
eastern edge of a patch of obscuration (Fig.~\ref{s1276cmco}c), which
corresponds to a peak in the CO emission, and the H{\sc ii} region
must therefore lie on the near side of the molecular complex.
This is confirmed by the fact that Fich et al. (\cite{fichha}) find
$V_{\rm lsr}$(H$_{\alpha}$) = $-98 \pm 0.4$~kms$^{-1}$, which indicates that
the ionized gas is moving towards us with respect to the molecular cloud.
The radio peak B (Fig.~\ref{s1276cmco}b) is slightly displaced to
the West with respect to the optical emission (Fig.~\ref{s1276cmco}c), which
indicates that this region is still partially obscured. At the radio peak
$\int T{\rm (CO)d}v \approx 41.3$~Kkms$^{-1}$, which translates
into $N{\rm (H_2)} \approx 7.8 \times 10^{21}$~cm$^{-2}$, corresponding to a
visual extinction $A_{\rm v} \approx$ 8~mag. Likewise, $\int T{\rm (CO)d}v$
seen projected on the optically visible part of the H{\sc ii} region 
corresponds to $A_{\rm v} \approx$ 2~mag, and probably  most of that CO lies 
behind the optically 
visible emission. The H{\sc ii} region therefore lies in a cavity in the 
southwestern molecular cloud; the cavity is open towards the East, where we 
can see the ionized gas optically, while the wall of the cavity
closest to us causes the rest of the H{\sc ii} region to be obscured by about
6~mag. The compact H{\sc ii} region is associated neither with a
temperature peak in Fig.~\ref{fig2_chan}, nor with a peak in $\int T{\rm d}v$ 
in Fig.~\ref{s1276cmco}. It also does not lie near the center of any of the
identified clumps. The clumps just to the SW and NE of the H{\sc ii}
region (resp. nrs. 11 and 14 in Table~\ref{s127cocl}) are at different
velocities (resp. --93.8 and --92.5~km$\,$s$^{-1}$). Because the ionized gas 
is breaking through the cloud, it has removed molecular gas from the 
line-of-sight. 
In Fig.~\ref{spectra_s127} we show CO spectra towards four offset positions:
that of the position of WB89~85 at (0\arcsec,0\arcsec), the compact H{\sc ii}
region to the SW ($-$16\arcsec,$-$22\arcsec), the CO peak near that
($-$20\arcsec,$-$30\arcsec), and the strong northern maximum 
(26\arcsec,50\arcsec).

\noindent
A comparison of the K-band emission with the optical image
shows there are about 40 objects that are visible in the NIR, but not on
the DSS. Considering the distance of this object, and its location in the 
outer Galaxy, it's unlikely that many, if any, of these K-band sources are 
reddened background stars. One of these objects lies
at the position of the peak of radio component A, and might be its exciting
source. These NIR data will be discussed in detail in a separate paper.
The IRAS source is seen projected on top of a small clump. It is  unclear 
whether there really is a YSO at this position
or whether the PSC position results from confusion due to nearby sources 
north and south.

\noindent

\begin{figure*}
  \resizebox{\hsize}{!}{\includegraphics{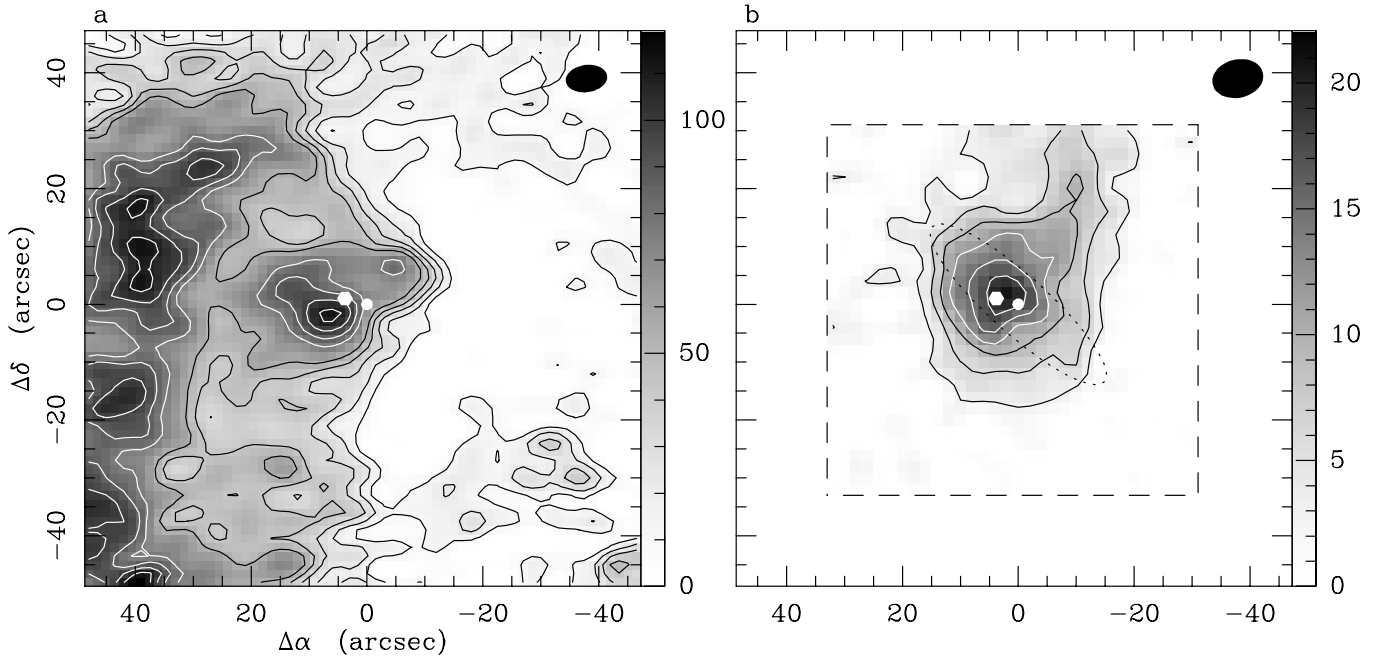}}
  \hfill
\caption{
High-resolution images of the WB89~380 region. The position of the compact
H{\sc {ii}} region (point source at 6~cm), detected with the VLA is indicated
by the hexagon. The IRAS source is shown as a filled circle.
{\bf a} The BIMA interferometer integrated ($-89.75 \leq V_{\rm lsr} \leq
-83.25$) CO map, with the contour levels: 11(11)~Kkms$^{-1}$; {\bf b} The 
integrated ($-90.55 \leq V_{\rm lsr} \leq -83.35$) CS distribution, with the
contour levels 3(3)~Kkms$^{-1}$. The error ellipse of the IRAS source is 
shown as well. There are no data in this panel outside the dashed box.}
\label{wb380cocs}
\end{figure*}

\smallskip\noindent
{\bf WB89~380} (IRAS~01045+6505). There is no association with an optical
H{\sc ii} region for this object. WB89 reported two velocity components
for this line of sight, of which the brighter one is at --88~km/s. More recent
$^{13}$CO and C$^{18}$O observations (Wouterloot \& Brand~\cite{wb96}) reveal
that the $^{12}$CO(1--0) is self absorbed, however, and that the central
velocity is actually --86~km/s. This leads to a kinematic galactocentric
distance of 16.6~kpc and a heliocentric distance of 10.3~kpc.
WB89~380 has a FIR luminosity of 1.0$\times$10$^5$~L$_\odot$, which suggests
the presence of an O7~V$_0$ star (or O8-8.5~V; Panagia~\cite{panagia}).
This object was observed at the VLA at 6, 3.6 and 2-cm by Rudolph et al.
(\cite{rudolph}); a point source was detected at the IRAS PSC position. From 
the continuum data, a spectral type of O7.5~V$_0$ (O8.5~V; Rudolph et 
al.~\cite{rudolph}) is derived, which is consistent with that deduced from 
the FIR luminosity.

\begin{figure*}
  \resizebox{12cm}{!}{\includegraphics{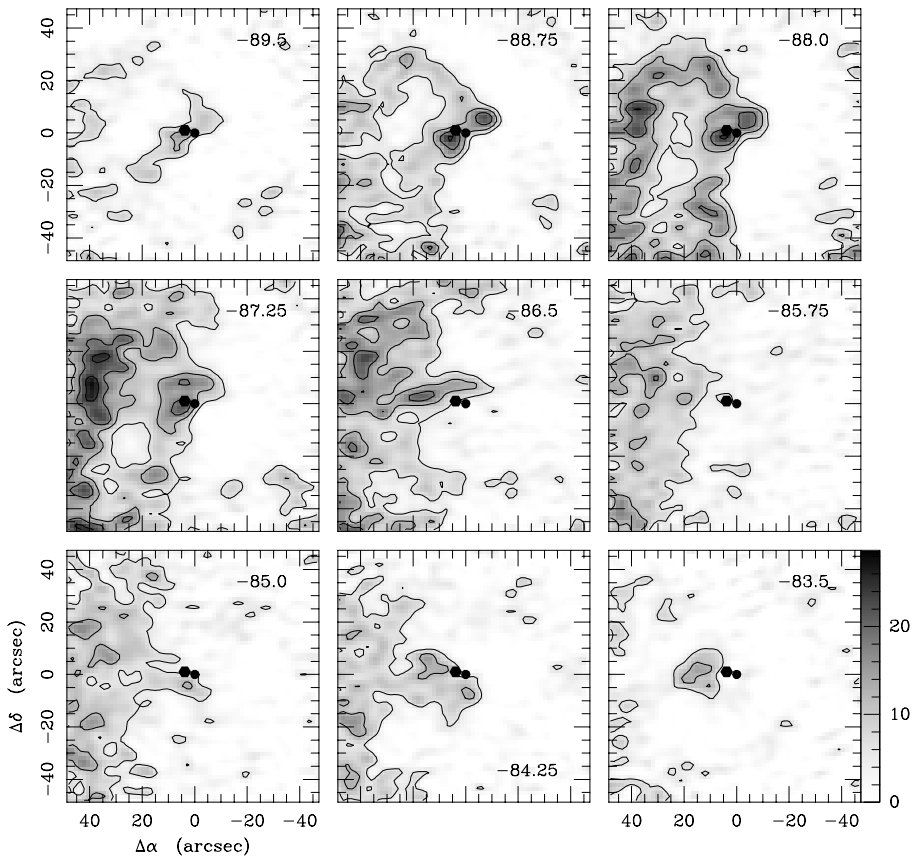}}
  \hfill
  \parbox[b]{55mm}{
  \caption{High-resolution $^{12}$CO(1--0) channel maps of the WB89~380 region
showing the velocity structure between --89.5 and --83.5~km$\,$s$^{-1}$.
Each panel shows the emission in a 0.75~kms$^{-1}$ wide interval, centered at 
the indicated velocity. Contour levels are 6(6)~Kkms$^{-1}$. The filled circle 
indicates the position of WB89~380, the hexagon that of the compact radio 
source.}
\label{bima380cochan}
}
\end{figure*}

\noindent
Fig.~\ref{kos380co} shows the single dish $^{12}$CO(2--1) map obtained
with the KOSMA~3-m telescope, from which we derive a mass $M_{\rm CO} =
3.5 \times 10^4$~M$_\odot$ ($X = 1.9 \times 10^{20}$, assuming a ratio
(2--1)/(1--0) of 0.85 (BW95), and $\eta_{\rm mb}=0.7$), and a radius
$r_{\rm eff}$ = 22.4~pc (see Table~\ref{cloudpar}).
In Fig.~\ref{kos380co} there is a strong maximum towards WB89~380, and a
weaker one towards WB89~379. Furthermore there is extended emission at a
level of about $T_{\rm A}^*$=1~K with line widths less than 2~km$\,$s$^{-1}$. 
The channel maps in Fig.~\ref{kos380cochan} show that most emission is at 
--88 to --86~km$\,$s$^{-1}$. Northwest of WB89~379 there is a separate 
component at about --86~km$\,$s$^{-1}$. Towards this IRAS source both 
components overlap (double lines are seen), while towards WB89~380 lines are 
much broader and profiles are asymmetric, and are affected by the 
self-absorption mentioned above. The highest 
temperatures occur south of this source. 

\smallskip\noindent
The BIMA observations are centered on the IRAS point source WB89~380, located 
near the strongest CO peak of the cloud. Fig.~\ref{wb380cocs}a shows the 
BIMA integrated CO emission, and Fig.~\ref{wb380cocs}b shows the same for the 
CS integrated emission. 

\begin{figure}
  \resizebox{\hsize}{!}{\includegraphics{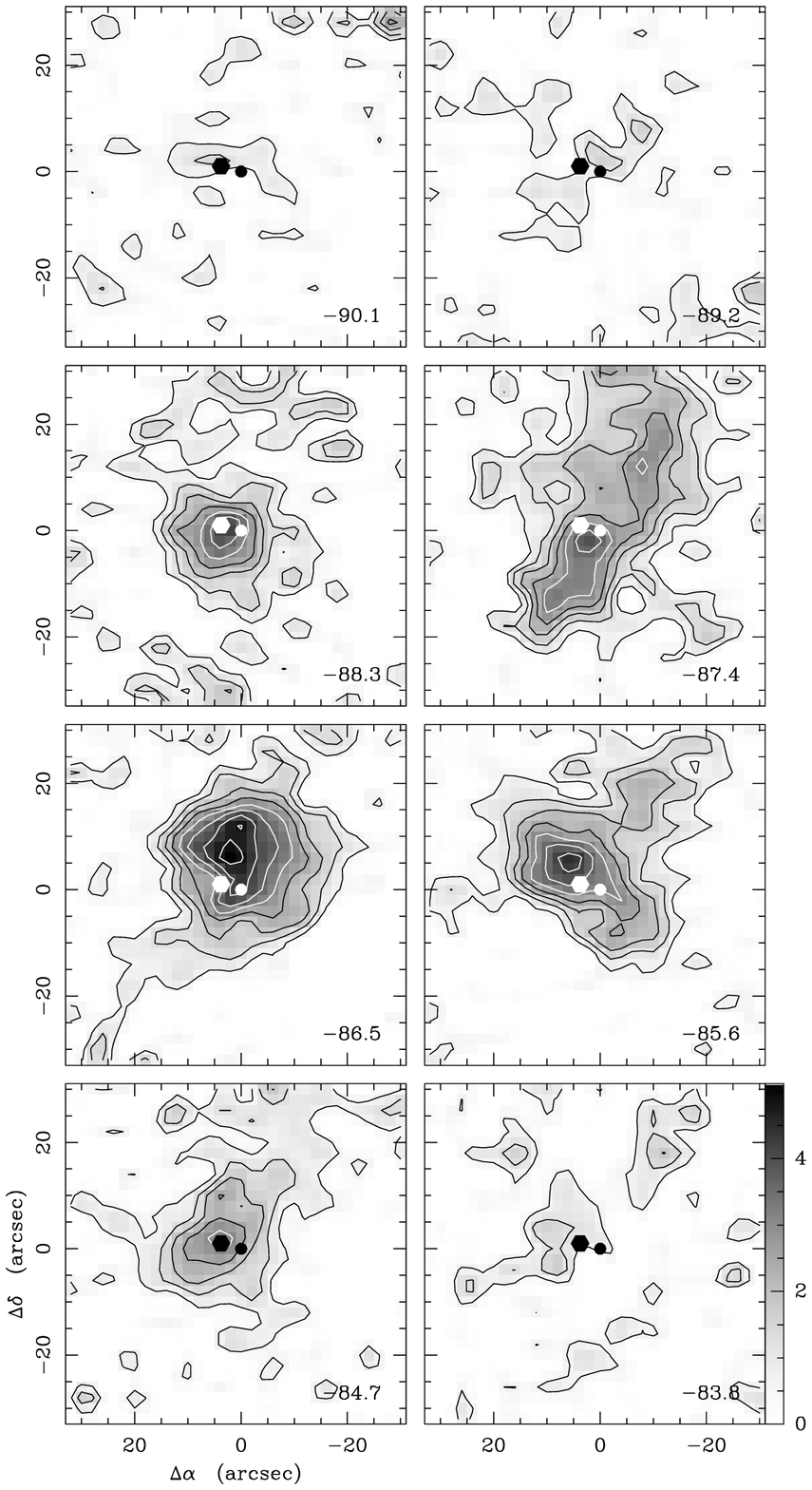}}
  \caption{High-resolution CS(2--1) channel maps of the WB89~380 region
showing the velocity structure between --90.1 and --83.8~km$\,$s$^{-1}$.
Each panel shows the emission in a 0.9~kms$^{-1}$ wide interval, centered at 
the indicated velocity. Contour levels are 0.6(0.6)~Kkms$^{-1}$. The filled 
circle indicates the position of WB89~380; the hexagon identifies the compact 
radio source.}
\label{bima380cschan}
\end{figure}

\begin{figure}
  \resizebox{\hsize}{!}{\includegraphics{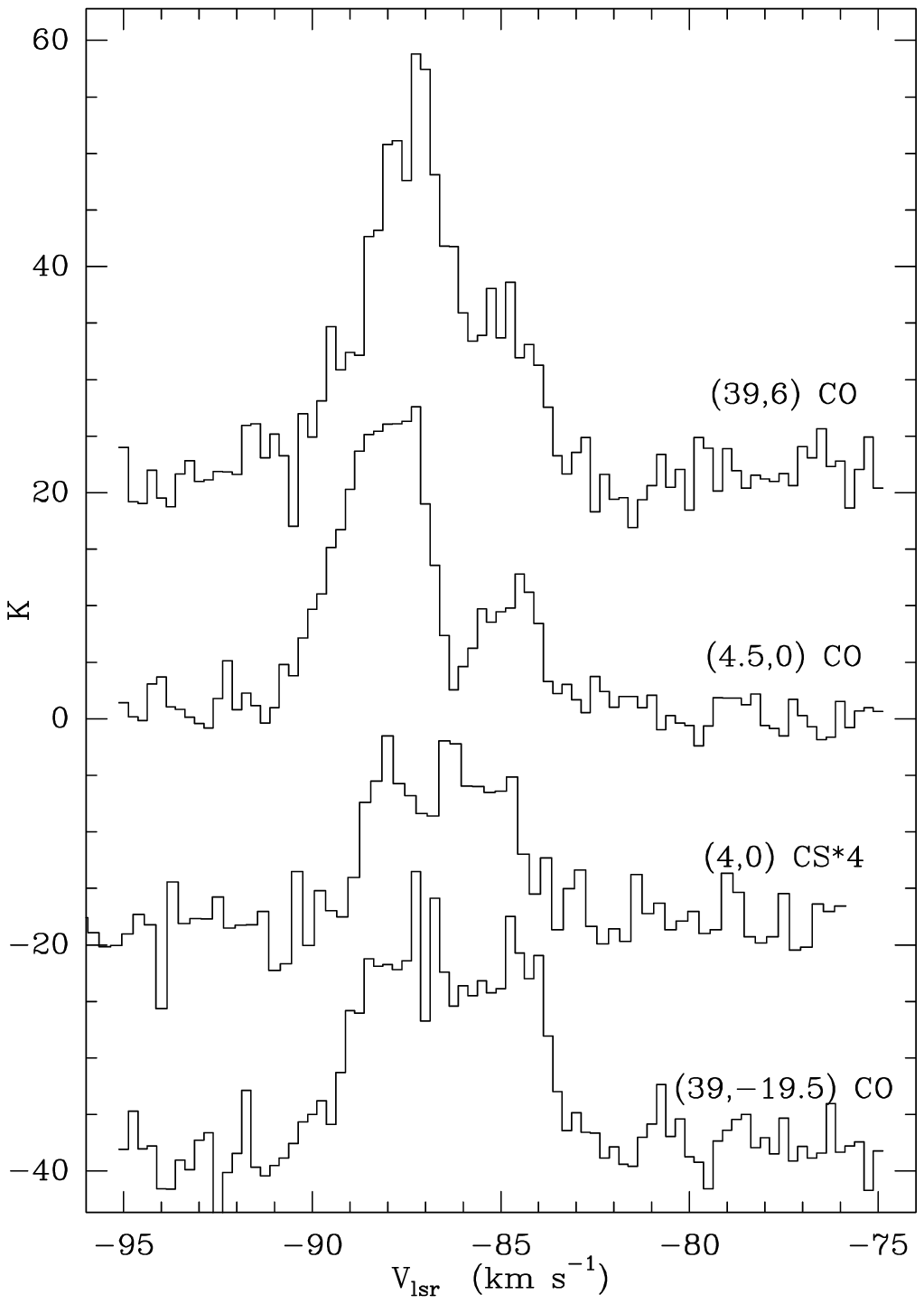}}
\caption{Spectra at three positions towards WB89~380 extracted from the CO and 
CS datacubes, averaged over 5$\times$5 pixels of 1\farcs5 (CO) or 2\arcsec\ 
(CS), centered at the indicated offsets from the map center. The absence of CO
emission at the line center (i.e. the location of the CS emission peak) is 
due to self absorption. }
\label{spectra_wb380}
\end{figure}

\noindent
The unresolved compact H{\sc ii} region lies at the edge of a peak
in the CO intensity map, but its position coincides almost exactly with the 
peak in the CS distribution, suggesting that the ionizing star(s) is (are) 
still embedded in the dense gas. This is confirmed by the detection of an 
H$_2$O maser towards WB89 380 (Wouterloot et al.~\cite{wbf}; {Comoretto et 
al.~\cite{comore}).
The velocity structure of the CO and CS emission is shown in
Figs.~\ref{bima380cochan} and \ref{bima380cschan}, respectively. There is
an indication of a ring-like structure in the CO line at
velocities between --89 and --87~km$\,$s$^{-1}$, perhaps caused by dynamical
interaction of earlier formed stars with the gas. 

\begin{figure*}
  \resizebox{12cm}{!}{\includegraphics{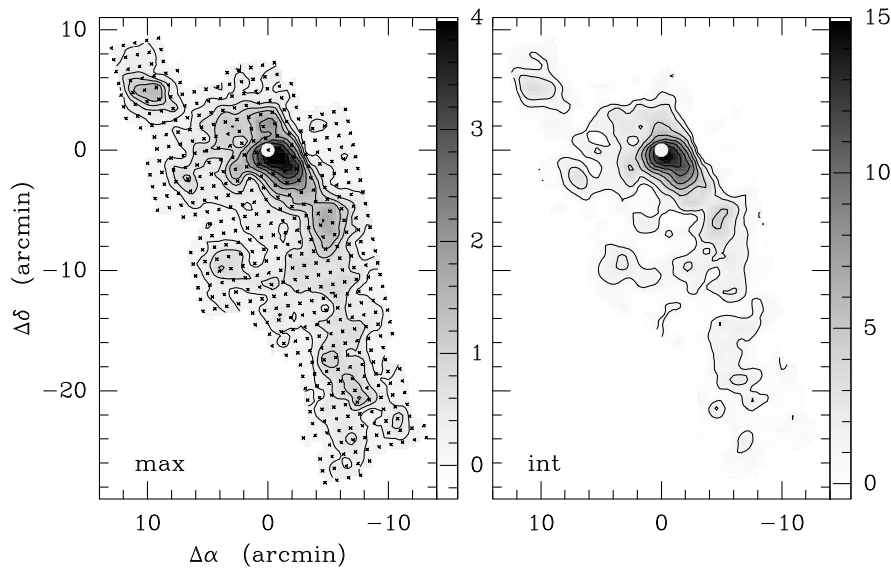}}
  \hfill
  \parbox[b]{55mm}{
\caption[ - ]{KOSMA $^{12}$CO(2--1) map of the cloud associated with WB89~399, 
showing the peak (left) and integrated CO intensity (right) from 
$-86 < V_{\rm lsr} < -77$~km$\,$s$^{-1}$. The contour levels are 0.3 (0.3) 1.2 
(0.6) 3.6~K(left) and 1(1)~K~km$\,$s$^{-1}$ (right). The circle at(0,0) 
indicates the location of the IRAS point source. The small crosses 
indicate the observed positions.} 
\label{kos399co}}
\end{figure*}

\noindent
In the CS
channel maps we note that while at the extreme velocities ($< -88$ and $>
-85$~kms$^{-1}$) the emission peaks at the location of the compact radio
source, it is offset from this position at the intermediate velocities,
where it also has a larger extent. This behaviour is reminiscent of that of 
an expanding or contracting shell around the radio source, and deserves study 
at higher angular resolution. 

\noindent
We can use the present data to estimate the age of the
H{\sc ii} region. Assuming the structure we see in the CS channel maps in
Fig.~\ref{bima380cschan} is in fact an expanding shell, i.e. gas being swept
up by the expanding H{\sc ii} region, the expansion velocity of the shell
$\approx 1.8$~kms$^{-1}$. The mass of the gas is $\sim$350~M$_{\odot}$,
which is the average between the LTE mass of the gas (assuming
$T_{\rm ex}=30$~K, and a [CS]/[H$_2$] abundance of 1.0$\times 10^{-8}$) and
the sum of the virial masses of the clumps (see Table~~\ref{380cscl}). Then
the kinetic energy of the shell $E_{\rm kin} \approx 1.1 \times 10^{46}$~erg. 
The central star is of type O7~V$_0$, has a luminosity of 10$^5$~L$_{\odot}$, 
of which a fraction of 0.35 (Panagia~\cite{panagia}) is available in the Lyman
continuum, i.e. $1.3 \times 10^{38}$~ergs$^{-1}$. The efficiency with which
this stellar energy is transformed into mechanical energy is about 0.1\%
(Spitzer~\cite{spitzer}). To arrive at the observed kinetic energy of the
shell thus requires $\sim 3000$~yrs. \hfill\break\noindent
The radius of the H{\sc ii} region is 0.65~pc (Rudolph et al.~\cite{rudolph}). 
An H{\sc ii} region around an O7~V$_0$ star reaches this
radius after $\sim 1.2-3.8 \times 10^3$~yrs (Spitzer~\cite{spitzer}), for 
initial densities of the embedding medium of $10^7 - 10^8$~cm$^{-3}$ (a
reasonable estimate, see also De Pree et al.~\cite{depree}). This calculation
is for the case of no absorption of the Lyman continuum photons by dust; with 
a fractional absorption by dust, it takes somewhat longer to reach a certain 
radius at the same initial density, but for the purpose of making a
rough estimate the difference is negligible. This time interval is
consistent with the 3000~yrs estimated above. In these considerations we
have ignored the contribution of stellar wind to the energy budget: with a
mass loss rate of $\sim 10^{-7.5}$~M$_{\odot}$yr$^{-1}$ and a wind velocity 
of $\sim 10^3$~kms$^{-1}$ (Garmany et al.~\cite{garmetal}) the available energy
$\approx 2.4 \times 10^{34}$~ergs$^{-1}$, four orders of magnitude less than
the energy available in Lyman continuum photons.

\noindent
Fig.~\ref{spectra_wb380} shows spectra for the CO and CS lines
derived from the interferometer data, by summing all spectra in a box of
7\farcs5$\times$7\farcs5 (CO) or 10\arcsec $\times$10\arcsec\ (CS) towards 
three offset positions. The CS spectrum peaks at --87~km/s, where the CO 
spectrum shows a dip, due to self-absorption. 
From these maps it is evident that the source is in an early stage of 
evolution with the young stellar object(s) still embedded in a high density 
clump. At 2 $\mu$m de Geus \& E. Lada (pers. comm.) have found a cluster of 
stars centered on the CS clump. WB89~380 is therefore in an earlier stage of 
evolution than WB89~85, where alongside a young region (the northern CS core 
and possibly the NIR objects, if they are embedded) we also find more evolved 
objects.

\begin{figure}
  \resizebox{\hsize}{!}{\includegraphics{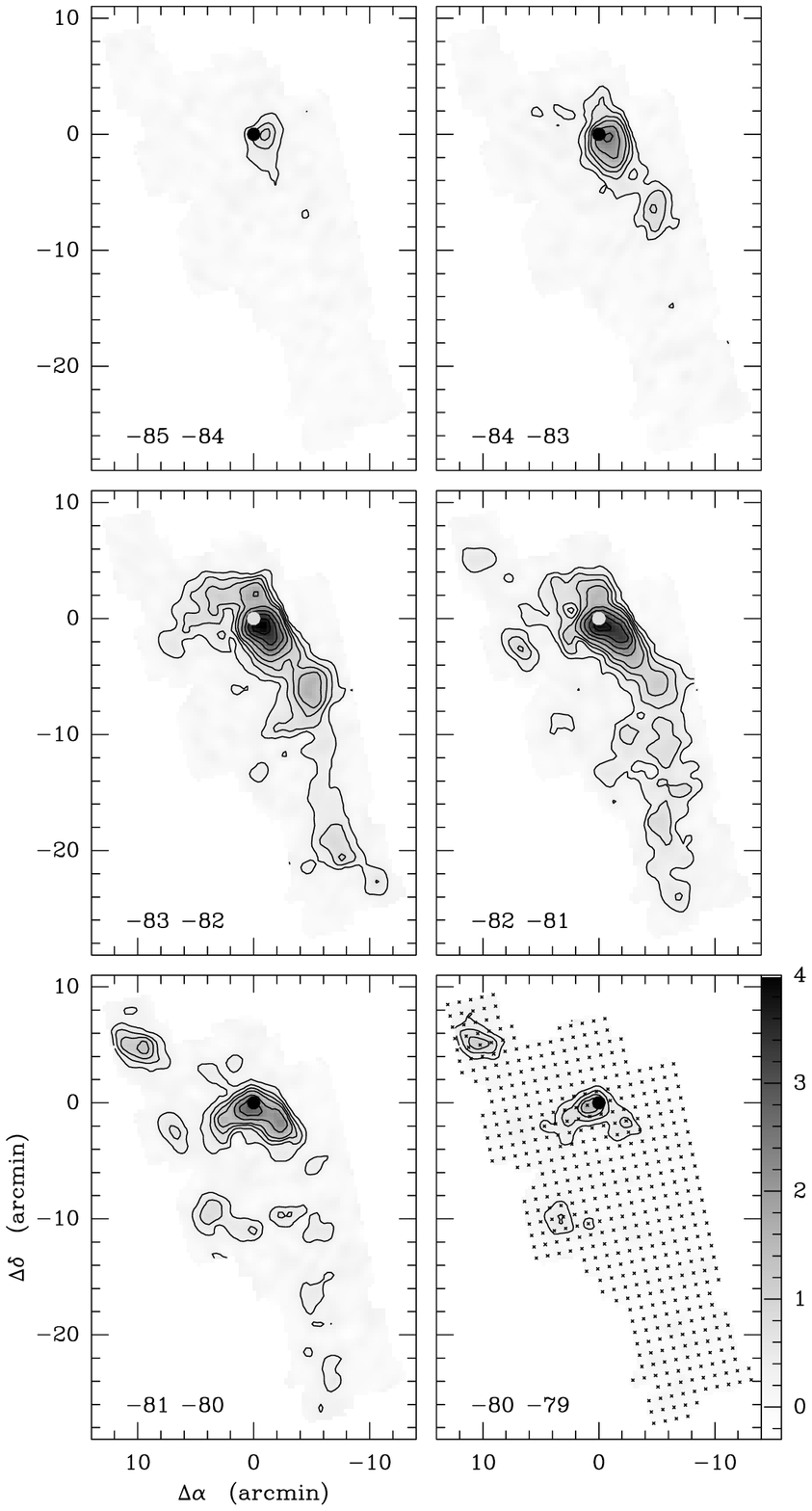}}
\caption[ - ]{KOSMA $^{12}$CO(2--1) observations of WB89~399, showing channel 
maps in intervals of 1~km$\,$s$^{-1}$. The contour levels are 0.3 (0.3) 1.2 
(0.6) 3.6~K~km$\,$s$^{-1}$. Symbols are as in Fig.~\ref{kos399co}.}
\label{kos399cochan}
\end{figure}

\smallskip\noindent
{\bf WB89~399} (IRAS~01420+6401). Towards WB89~399 an (optically visible)
resolved H{\sc ii} was detected with the VLA (see Rudolph et
al.~\cite{rudolph}). The radio data suggest excitation by an O9.5~V$_0$ star,
while the FIR luminosity of 4.3$\times 10^4$~L$_\odot$ implies a spectral type
of O9-9.5~V$_0$ (or B0~V; Panagia~\cite{panagia}). The region was originally
mapped in $^{12}$CO(1--0) and, incompletely, in $^{12}$CO(2--1) by
Brand \& Wouterloot (\cite{brandw1}). We present in Fig.~\ref{kos399co} a new,
complete $^{12}$CO(2--1) map of this region, made with KOSMA; due to lack of
time, no BIMA observations were made. There is a strong maximum slightly
southwest of WB89~399, located in a ridge of emission. In addition there are
more isolated and weaker cloud components, in general with small line widths.
Fig.~\ref{kos399cochan} shows the velocity distribution. The strongest
emission is at --83 to --81~km$\,$s$^{-1}$ with some weaker components at
other velocities. The mass derived from the CO(2--1) data, assuming an
intensity ratio of (2--1)/(1--0) of 0.85, $\eta_{\rm mb}=0.7$, and $X =
1.9 \times 10^{20}$, is $M_{\rm CO} = 2.6 \times 10^4$~M$_{\odot}$; the
effective radius $r_{\rm eff}$ = 21.3~pc (see Table~\ref{cloudpar}).

\begin{figure}
  \resizebox{\hsize}{!}{\includegraphics{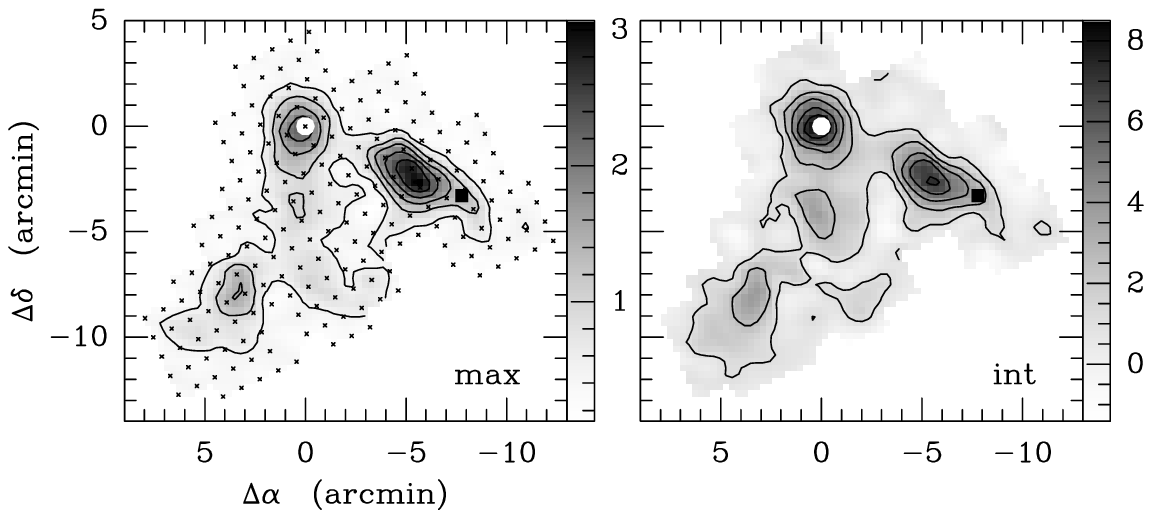}}
\caption[ - ]{KOSMA $^{12}$CO(2--1) map of WB89~437, showing the peak (left)
and integrated CO intensity (right) from 
$-80 < V_{\rm lsr} < -68$~km$\,$s$^{-1}$. The lowest contour level and contour 
step are 0.5 and 0.5K (left) and 1.0 and 1.5~K~km$\,$s$^{-1}$ (right). The 
triangles show the locations of two IRAS point sources that were included in 
Wouterloot \& Brand (\cite{wb89}): WB89~436 (IRAS 02383+6241; square) and 
WB89~437 (circle). The small crosses indicate the observed positions. The 
interferometer observations are centered on the position of WB89~437.}
\label{kos437co}
\end{figure}

\smallskip\noindent
{\bf WB89~437} (IRAS 02395+6244). This object has a FIR luminosity of
7.1$\times$10$^4$~L$_\odot$, which indicates the presence of an O8~V$_0$
star (or O9-9.5~V; Panagia~\cite{panagia}). In the radio continuum
observations, however, WB89~437 was not detected, putting an upper limit to
the spectral type of a (single) embedded object of B1 (Rudolph et
al.~\cite{rudolph}). These authors considered various scenarios
to explain the discrepancy between the radio continuum and FIR observations,
such as the presence of a very rich cluster of non-ionizing main-sequence
stars of spectral type later than B1, large enough to produce the
$L_{\rm fir}$ of 7.1$\times 10^4$~L$_\odot$, or the presence of a large amount
of dust within the H{\sc ii} region, that would have to absorb $\sim$90\% of
the UV radiation.
Other possibilities are that an H{\sc ii} region is present, but that it is
very young and still optically thick. It is also possible that $L_{\rm fir}$ 
is generated by infall; a small amount of infalling matter would be enough to 
quench the forming of an H{\sc ii} region around the embedded massive object 
(Molinari et al.~\cite{molinari}; Walmsley~\cite{walmsley}).

\begin{figure}
  \resizebox{\hsize}{!}{\includegraphics{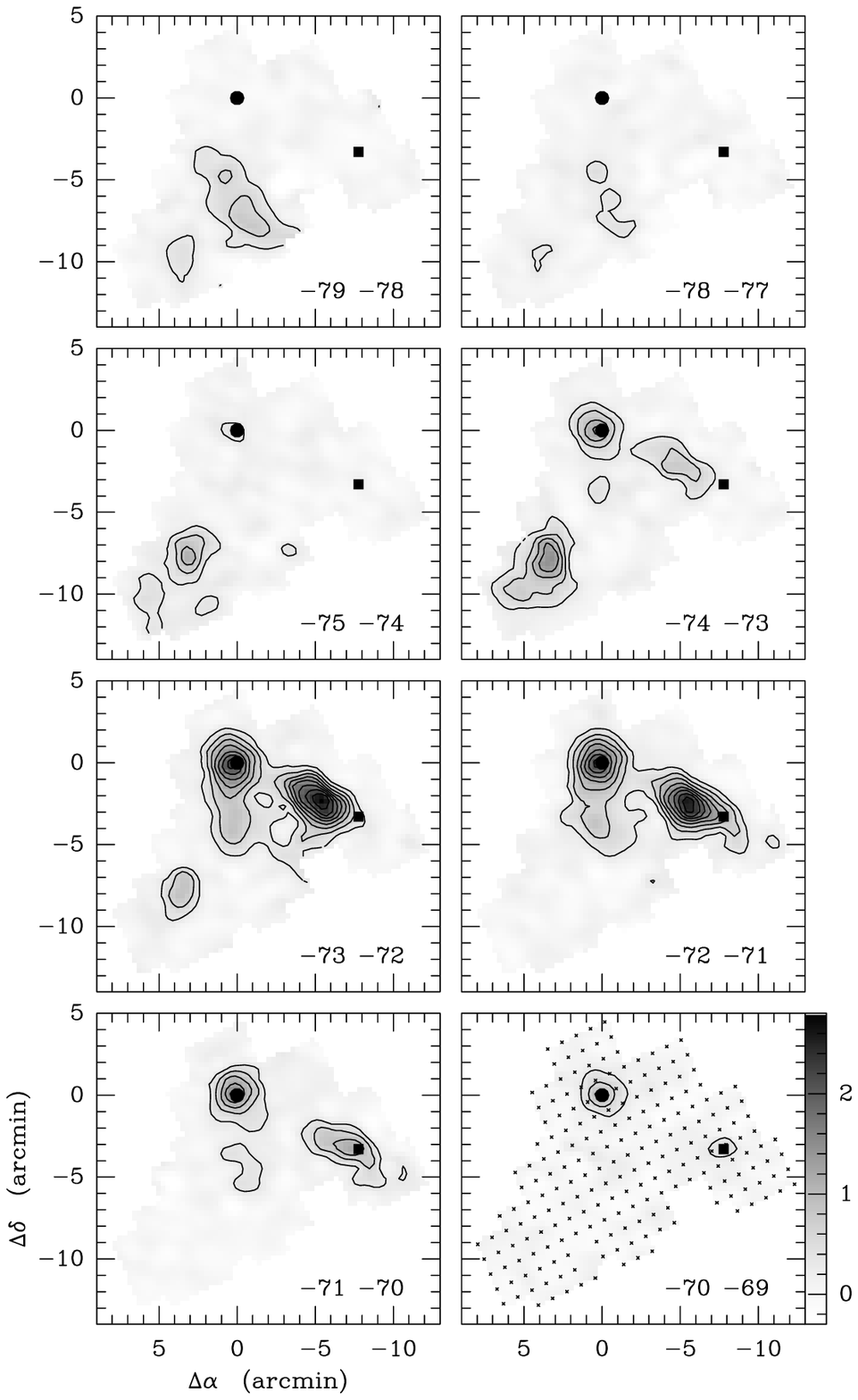}}
\caption[ - ]{KOSMA $^{12}$CO(2--1) observations of WB89~437, showing channel 
maps in intervals of 1~km$\,$s$^{-1}$. The lowest contour level and contour 
step are 0.3~K~km$\,$s$^{-1}$. Symbols are as in Fig.~\ref{kos437co}.}
\label{kos437cochan}
\end{figure}

\noindent
Fig.~\ref{kos437co} shows the KOSMA $^{12}$CO(2--1) map of WB89~437.
With the same assumptions as before, we derive a mass $M_{\rm CO} = 9.4 \times
10^3$~M$_\odot$, and a radius $r_{\rm eff}$ = 13.3~pc (see 
Table~\ref{cloudpar}). There are two main clumps, near WB89~436 and 437 
respectively, and several cloud components with weaker emission. The optically 
visible WB89~436 (see Rudolph et al.~\cite{rudolph}) is displaced from the 
center of the associated CO cloud, whereas WB89~437 is deeply embedded and 
shows strong outflow emission (Wouterloot \& Brand~\cite{wb96}).
Channel maps of the $^{12}$CO(2--1) emission are shown in 
Fig.~\ref{kos437cochan}. The main clouds are at a velocity of $-73$ to 
$-71$~km$\,$s$^{-1}$, but the components in the southeastern part show 
emission at velocities down to --79~km$\,$s$^{-1}$.

\begin{figure*}
  \resizebox{\hsize}{!}{\includegraphics{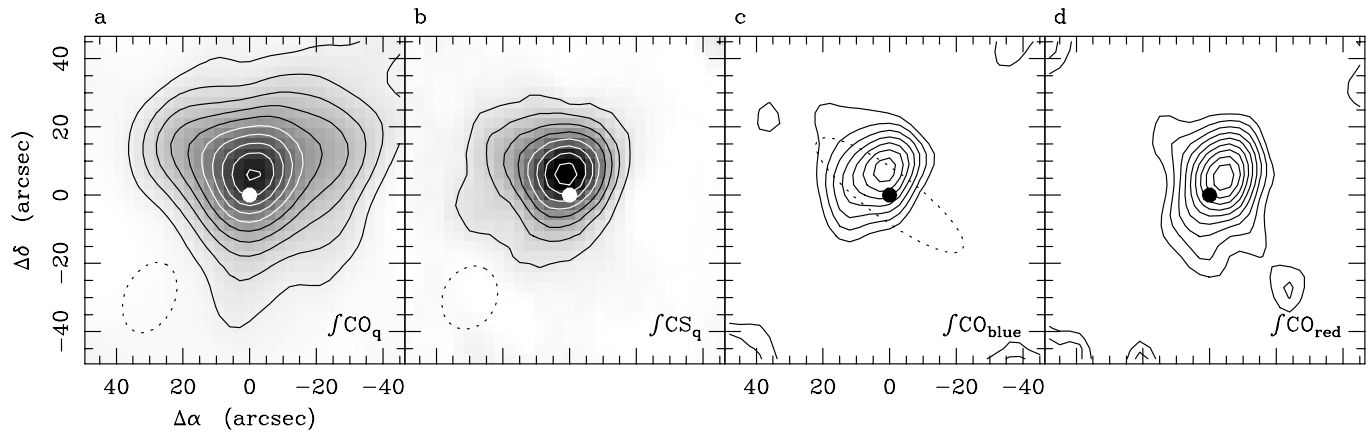}}
\caption{Results of the mm-interferometer measurements of WB89~437. The circle
indicates the IRAS PSC position; the dotted ellipse in panels a and b
represents the beam; the dotted contour in panel c is IRAS error ellipse. No 
primary-beam correction has been applied.
{\bf a} Integrated CO emission from $-$75 to $-$67.5~km$\,$s$^{-1}$. Lowest 
contour level and step are 17~Kkms$^{-1}$. The range of the grey scale is -10 
(white) to 185~Kkms$^{-1}$ (black).
{\bf b} Image of integrated CS emission from from $-74$ to
$-68.9$~km$\,$s$^{-1}$. Contour levels are 1.8 to 14.4~Kkms$^{-1}$ in steps of
1.8~Kkms$^{-1}$. The range of the greyscale is -1 (white) to 14~Kkms$^{-1}$
(black).
{\bf c,d} Contour plot of the CO emission, integrated over velocity
intervals covering the blue (c) and red wing (d). The integration intervals
are  $-81<V_{\rm lsr}<-78~$km$\,$s$^{-1}$ and
$-66.5<V_{\rm lsr}<-63.5$~km$\,$s$^{-1}$. The contour levels are
1.4(1.4)~Kkms$^{-1}$.}
\label{wb437csco}
\end{figure*}

\noindent
The BIMA observations were centered on the IRAS point source. 
Fig.~\ref{wb437csco}a and b show the distribution of the integrated CO and CS 
emission, respectively.
Unfortunately only C-array interferometer observations were obtained for 
this source in both CO and CS, which results in maps of relatively low 
resolution (18\arcsec =0.8~pc). 
Both CO and CS peak near the position of the IRAS point source, and 
the observations reveal a compact clump in both molecules.
Near-IR observations of WB89~437 have led to the detection of an
embedded young cluster centered on the dense core of the molecular gas (de Geus
\& E. Lada, pers. comm.). Fig.~\ref{wb437csco}c and d show two 
velocity-integrated maps, covering the velocity ranges of the blue and red 
wings of the profile, respectively;  the emission centers are offset by 
6\arcsec.  Spectra at the peak of the CO and CS emission at 
(0\arcsec,6\arcsec) are shown in Fig.~\ref{spectra_wb437}. The presence of 
strong wing emission is evident from the CO spectrum, confirming the 
interpretation that the embedded cluster is producing outflowing gas.
Unfortunately the available velocity range is too small to cover all outflow 
gas (cf. Fig.~\ref{spectra_wb437}), and therefore we cannot derive the outflow 
parameters. The IRAM CS observations (not shown) also show this outflowing gas.
The early evolutionary state of this source is also confirmed by the detection 
of an H$_2$O maser towards this source (Zuckerman \& Lo~\cite{zuck}; 
Wouterloot et al.~\cite{wbf}). 

\begin{figure}
  \resizebox{\hsize}{!}{\includegraphics{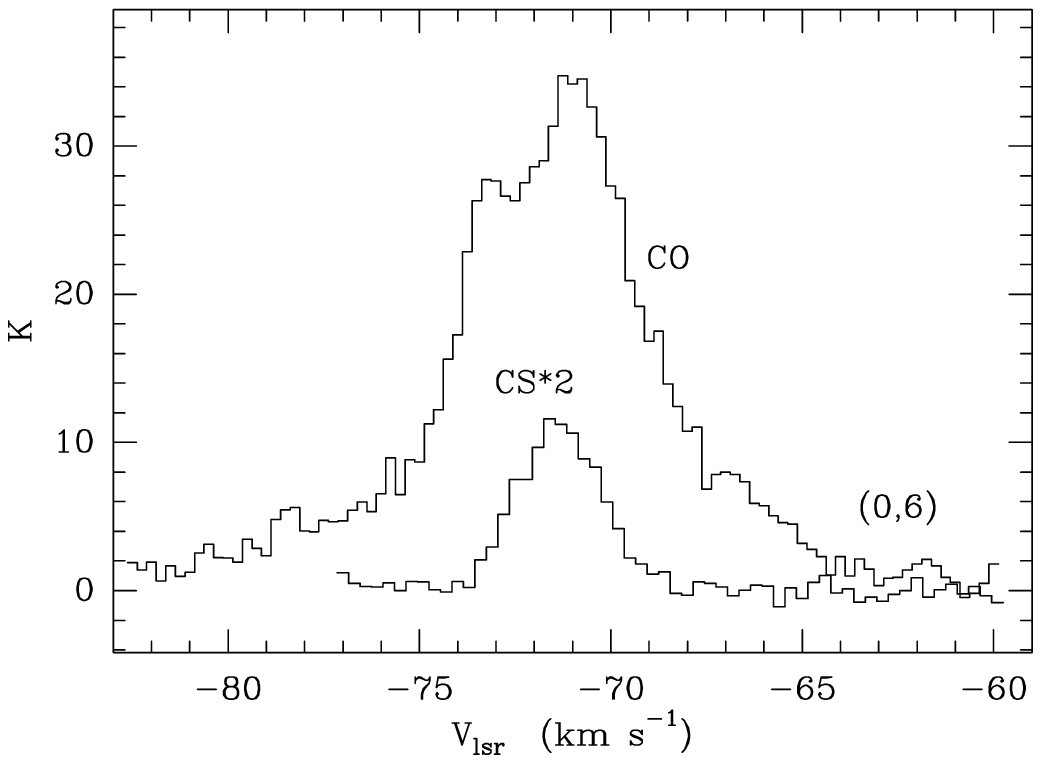}}
\caption{Spectra towards the peak of the integrated CO and CS emission of 
WB89~437, extracted from the BIMA datacubes, averaged over 5$\times$5 pixels 
of 3\arcsec\ centered at the indicated offset (in arcsec) from the map center.}
\label{spectra_wb437}
\end{figure}

\section{Clump Analysis \label{clumpana}}

The observations have been analysed using the three-dimensional clump detection
and analysis program CLUMPFIND described by Williams et al. (\cite{williams}).
The clump analysis has been carried out on both the CO and the CS data. 
Table~\ref{s127cocl} lists the properties of the CO clumps in WB89~85, the 
clump properties for WB89~380 are listed in Table~\ref{380cocl} (from CO) and 
\ref{380cscl} (from CS), and Table~\ref{437cocscl} lists the clump properties 
for WB89~437 (both CO and CS). Note that the tables list only the 
{\sl resolved} clumps, i.e. those for which an accurate value of the radius 
could be derived. Most of the clump parameters are self-explanatory,
but a few require additional comments. 

\noindent
Col.~7 gives the effective radius $r_{\rm eff}$ of the clumps, which was
derived from the measured effective radius at the lowest usable intensity
level of the map (which we put at 2.5~$\sigma_{rms}$):
$r_{\rm eff} = \sqrt{r_{\rm eff,obs}^2 - r_{\rm beam}^2}$, where
$r_{\rm beam}$ is the beam size. $r_{\rm eff,obs} = \sqrt{A/\pi}$ has been
obtained by measuring the area ($A$) of the total extent of the clump when
integrated over its full velocity range.
At this low intensity level (relative to the clump maximum), the beam will have
a size which differs from the usually reported full-width at half maximum
(FWHM) beam size. Assuming the beam is Gaussian, the appropriate beam size can
be calculated for each clump from the ratio of the peak temperature
($T_{\rm peak}$) of the clump to the actual temperature ($T$) at which the
size is measured, and expressed in terms of the $FWHM$: $r_{\rm beam} = 0.6
\sqrt{\ln T_{\rm peak} - \ln T} \times FWHM_{\rm beam}$.  

\noindent
The FWHM linewidths of the clumps, $\Delta v$, are listed in column~8 and were
calculated from the formal one-dimensional velocity dispersion $\sigma$
through $\Delta v^2$=8ln2~$\sigma^2$.

\noindent
The CO luminosity, $L_{\rm CO} ($K$\,$km$\,$s$^{-1}$ pc$^2$)
is calculated by adding the signals of all pixels that are assigned to a
particular clump, and multiplying with the channel width in km$\,$s$^{-1}$ and
the pixel area in pc$^2$. The mass of the clump is then determined from
$M_{\rm CO}$/M$_\odot$=5.0\ ($X$/2.3)\ $L_{\rm CO}$ (including a correction of
1.36 for He), where $X$ is the conversion factor from CO integrated intensity
to molecular hydrogen column density (in units of $\rm 10^{20}
(K km s^{-1})^{-1} cm^{-2}$). We use $X$=1.9, the value derived by Strong \&
Mattox (\cite{strongmat});
hence $M_{\rm CO}$/M$_\odot$=4.13\ $L_{\rm CO}$ (Col.~9).
The lowest, highest $M_{\rm CO}$ found are 3.3, 349~M$_{\odot}$ in WB89~85,
2.0, 62.4~M$_{\odot}$ in WB89~380, and 9.7, 480~M$_{\odot}$ in WB89~437. 
CLUMPFIND managed to allocate virtually all emission in the maps to clumps:
99.4\% for WB89~85, 94.9\% for WB89~380, and 98.2\% for WB89~437. However, the
original lists contain a large number of clumps that are unresolved, lie
(partially) over the edge of the mapped region, or look unreal upon visual
inspection. We have therefore trimmed the lists created by CLUMPFIND to
include only real, resolved clumps. In the end, 67\% of the mass in all
clumps found is in such clumps for WB89~85; for WB89~380 this is 74\%, and
for WB89~437 73\%.  

\noindent
The virial mass is given in Col.~10 (Col.~9 in Table~\ref{380cscl}). For a
spherical clump with radius $r$, a one-dimensional
velocity dispersion, $\sigma_v$, and a density distribution of $\rho (r)
\propto r^{-n}$ (MacLaren et al.~\cite{maclaren}):
$$M_{\rm vir} = \left({5 - 2n \over 3 - n} \right) \,
{3\sigma_v^2 r \over {\rm G}}.$$
This equation ignores contributions like magnetic fields, internal heating,
and external pressure. With $n$=2, we obtain the following expression for the 
virial mass: $M_{\rm vir} = 126\ r\ \Delta v^2.$ 

\subsection{Local comparison sample \label{locsamp}}

In the following we shall discuss the physical parameters of the clumps, and
compare them to those of a typical nearby GMC. As comparison objects we have
chosen the \object{Rosette Molecular Cloud} (RMC), and the 
\object{Orion~B South} cloud (containing NGC~2023 and 2024). 

\smallskip\noindent
The RMC data are those used in Williams et al. (\cite{willblitz});
Jonathan Williams kindly made the original data cube available.
The resolution ($\approx 0.8$~pc) and sampling ($\approx 0.7$~pc) of these
$^{13}$CO(1--0) observations is lower than for the outer Galaxy clouds.
We ran CLUMPFIND on these data, after which clump
parameters were calculated in the same way as described above for the three
outer Galaxy clouds. The clump properties are listed in Table~\ref{rmclowcocl}.
In this case, about 95\% of all mass was distributed in
clumps. After visual inspection of all clumps, 46 were judged to be real and
resolved (these contain $\sim$82\% of the mass in the original list). We note
that we find fewer clumps than do Williams et al. (\cite{willblitz}), which is
mostly due to the fact that their list contains several clumps that lie on
or partially over the edge of the mapped area, or that do not appear to be
real. For the mass calculation of those clumps we find in common with
Williams et al. we used the $T_{\rm ex}$ they give in their clump table,
allowing the application of an optical depth correction to the $^{13}$CO
column density $N(^{13}$CO): 
$$N(^{13}{\rm CO}) = f(T_{ex}) \times \left({\tau \over 1 - exp(-\tau)}
\right) \, \int T_{\rm R}^{\ast} {\rm d}v,$$
where $f(T_{\rm ex}$) is of the order of 1.1$\times 10^{15}$ for
$T_{\rm ex}$ between about 5 and 20~K.
The column density of H$_2$ was derived using an abundance
ratio [H$_2$]/[$^{13}$CO] = 5.0 $\times 10^5$, appropriate for local clouds
(Dickman~\cite{dickman}; Dickman \& Herbst~\cite{dickherbst}).
For the remaining clumps without $T_{\rm ex}$ data, an average factor was
used, based on the $f(T_{\rm ex}$) of the other clumps.  

\noindent
To prevent a comparison being biased by the lower resolution of these RMC
data, we have also used IRAM 30-m $^{13}$CO(2--1) data of three regions in
the RMC, as published by Schneider et al. (\cite{schneider}). These data were
taken with an 11\arcsec\ beam on a 15\arcsec\ (0.12~pc) grid, and were made
available by Nicola Schneider. We derived masses in the same way as for
the lower-resolution Williams et al. data.
Following Schneider et al. (\cite{schneider}) we assumed $T_{\rm ex}$=30~K,
yielding $f(T_{\rm ex}$)= 0.58 $\times 10^{15}$; a change in $T_{\rm ex}$ of
$\pm$10~K would change $f(T_{\rm ex}$) by 17\%.
The clump properties are listed in Table~\ref{rmchighcocl}, which combines 
the three regions Schneider et al. (\cite{schneider}) refer to as
``Monoceros Ridge'', ``The Central Part'', and ``Extended Ridge'',
respectively.

\smallskip\noindent
The Orion~B South data set has been described by Kramer et al.
(\cite{kramer96}; \cite{kramer}).
The original $^{13}$CO(2--1) data cube was made available to us by
Carsten Kramer. The advantage of this data set is its high spatial resolution;
the data were originally taken on a one beam FWHM ($\approx$0.3~pc) grid, and
later resampled to a 1\arcmin\ ($\approx$0.14~pc) grid.
CLUMPFIND was used to create a list of clumps that contain 99.3\%
of the mass in the map. After visual inspection we were left with a list of
33 clumps that were resolved, and which contain 87\% of the mass in the
original list. Masses were calculated as for the RMC, assuming 
$T_{\rm ex}$=25~K. All other parameters were derived as for the outer Galaxy 
clouds. The clump properties are listed in Table~\ref{oricocl}.

\begin{figure*}
\resizebox{12cm}{!}{\includegraphics{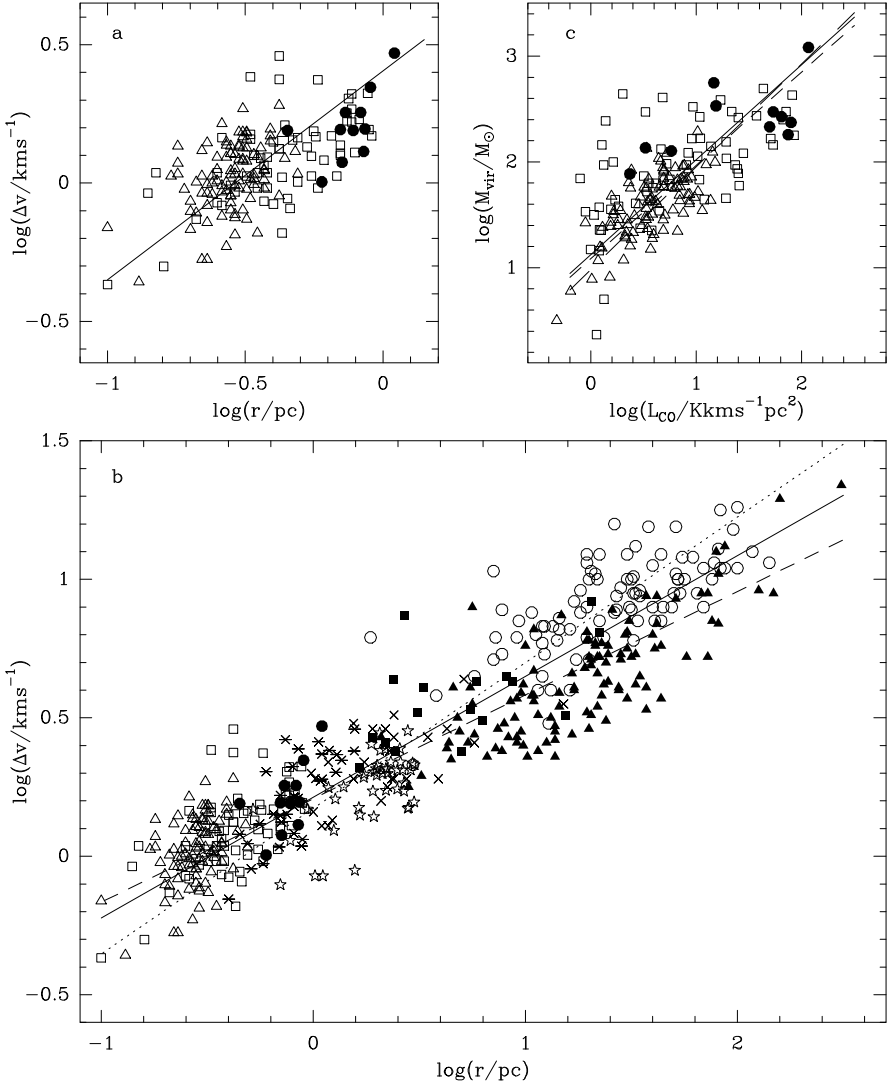}}
\hfill
\parbox[b]{55mm}{
\caption{{\bf a}\ Size-linewidth plot for the clumps found in the CO datasets
of WB89~85 (open squares), WB89~380 (open triangles) and WB89~437 (filled 
circles) plotted together. The drawn line indicates the linear least squares 
fit to the data.\hfill\break\noindent
{\bf b}\ As {\bf a}, but with additional data: inner Galaxy (open circles) and 
outer Galaxy (filled triangles) from BW95;
Rosetta nebula (lower-resol. data; stars) and Orion (asterisks) from this 
paper; Taurus (filled squares) from Ungerechts \& Thaddeus~\cite{unger}; 
High-latitude clouds (crosses) from Heithausen et al.~\cite{heith}. The drawn 
line is a least-squares bisector fit through all datapoints; 
the dashed line is a fit to outer Galaxy and BIMA data; the dotted line is a
fit to the inner Galaxy and local cloud data (Taurus, RMC, HLCs, Orion).
\hfill\break\noindent
{\bf c}\ Plot of the virial mass versus the CO luminosity for clumps in all 
three objects. The full line is the linear least squares fit to our data; 
the short-dashed line is the fit based on the BW95 data for inner Galaxy 
clouds; the long-dashed line (practically invisible, because coincident with 
the drawn line) is the fit to BW95 data for outer Galaxy clouds.}
\label{sizlinall}}
\end{figure*}

\section{Discussion \label{discuss}}

\subsection{CO clump properties}

\subsubsection{Size-linewidth relation\label{sizlinco}}

Fig.~\ref{sizlinall}a shows the size-linewidth plot of the CO-clumps in all
three sources.
The data cover a narrow range in size, and a weak correlation between these two
parameters is revealed. The formal linear-least-squares fit to the data gives:

\begin{equation}
{\rm \log} \Delta v = (0.41\pm0.02) + (0.75\pm0.04) \log r
\end{equation}

\noindent
(corr. coeff. 0.61; for all fits in this paper we used the bisector
fit: see e.g. Isobe et al.~\cite{isobe}). This is considerably steeper than
the size-linewidth relations found for samples of 
molecular clouds (see e.g. BW95; Blitz~\cite{blitz} for an overview), 
and is probably due to the small range in radius. A similar effect was noted 
by BW95, who from a fit to 112 outer Galaxy clouds found a slope of 
0.53$\pm$0.03 (corr. coeff. 0.79), while the individual datasets of which 
their total sample was composed, gave slopes between 0.39$\pm$0.05 and 
0.79$\pm$0.13.
To illustrate this point, Fig.~\ref{sizlinall}b shows the data from
Fig.~\ref{sizlinall}a in the context of a larger data set; the data are 
essentially from BW95, augmented by additional data from the literature (see 
caption) to bridge the gap in radius between the present data and those of 
BW95. It is clear that the present data are consistent with the range defined 
by the fits made to inner- and outer Galaxy data. A fit through all data 
points in this figure leads to

\begin{equation}
\log \Delta v = (0.21\pm0.01) + (0.44\pm0.01) \log r
\end{equation}

\noindent
(corr. coeff. 0.93).

\noindent
In Fig.~\ref{sizlinall}b we have mixed data for whole clouds with
data for clumps within clouds (Taurus, RMC). In fact, size-linewidth
relations derived for clumps show a range of slopes similar to what is found
for samples of clouds. For our comparison sample we find 0.74$\pm$0.09
(corr. coeff. 0.61) and 1.03$\pm$0.09 (corr. coeff. 0.76) for the RMC and
Orion, respectively. Carr (\cite{carr}) derives 0.24$\pm$0.06 for $^{13}$CO
clumps in Cep~OB3, while Loren (\cite{loren}) finds that $\Delta v$ is 
independent of size for $^{13}$CO clumps in $\rho$Oph. And finally, Lada et 
al. (\cite{lada}), from CS data, find that the size-linewidth relation 
depends on clump definition. They, like others before them (Issa et 
al.~\cite{issa}; Combes~\cite{combes}) cast doubt on whether this relation 
really reflects the dynamical state of molecular clouds.

\smallskip\noindent
Fig.~\ref{sizlinall}c shows the virial mass of the clumps as a function of
the CO luminosity. The drawn line is a fit to these data, while the dashed
lines are fits to data for inner- and outer Galaxy clouds (BW95; see
caption). A good correlation is found between the virial mass and the
CO luminosity:

\begin{equation}
\log M_{\rm vir} = (1.13\pm0.04) + (0.90\pm0.05) \log L_{\rm CO}
\end{equation}

\noindent
(corr. coeff. 0.70), virtually identical to the fits obtained for inner- and
outer Galaxy clouds as a whole (BW95). As was shown by BW95, the lack of an
offset between inner- and outer Galaxy clouds in a diagram of $\log
L_{\rm CO}$ and $\log \Delta v$ (which is similar to what is shown in
Fig.~\ref{sizlinall}c) implies that there is no indication of a change in
$X$ (= $N{\rm (H_2)}/W_{\rm CO}$) between inner- and outer Galaxy.
Fig.~\ref{sizlinall}c shows that what applies to entire clouds also holds
for clumps. We note however that BW95 also concluded that a $\log
L_{\rm CO}$ vs. $\log \Delta v$ (or $\log M_{\rm vir}$) diagram 
may not be the correct instrument from which to actually derive a value of 
$X$. 

\begin{figure}
  \resizebox{\hsize}{!}{\includegraphics{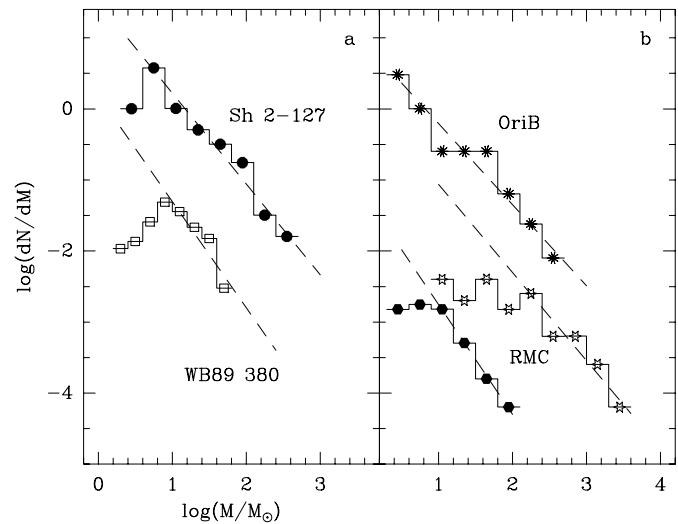}}
\caption{Plot of the clump mass distribution: log(d$N$/d$M$)\ vs.\ log$M$
{\bf a}\ WB89~85 and WB89~380 (shifted down by --2.0). The dashed
lines indicate linear least squares fit to the data above 4 and 6~M$_\odot$,
respectively for WB89~85 and WB89~380. {\bf b}\ as {\bf a}, for Orion, 
RMC (low-resolution data; shifted down by --1.5), and RMC (high-res. data;
shifted down by --3). The dashed lines are fits to the data above 2, 126, and 
8~M$_\odot$ respectively, for Orion, RMC (low-res.), and RMC (high-res.).}
\label{histodndm}
\end{figure}

\subsubsection{Mass distribution \label{massdist}}

An important parameter describing clumps in molecular clouds is the shape of
their mass distribution. Many different studies of nearby (see e.g.
Combes~\cite{combes}) and distant (BW95) molecular clouds have found very
similar mass distribution slopes between $-1.4$ and $-1.8$ in a plot of
log(d$N$/d$M$)\ vs.\ log$M$. Interestingly enough, the same slopes are
found for the mass distribution of clumps inside clouds (e.g. Kramer et al.
(\cite{kramer}), who studied the clumps in 7 clouds, with a range in masses
and resolution, and found slopes between $-1.6$ and $-1.8$).
Fig.~\ref{histodndm} shows the mass distribution for the clumps in WB89~85
and WB89~380, as well as for the clumps in the comparison clouds RMC
(lower-resolution data) and Orion~B South. For WB89~85 a fit to the data
for $M>4.0$~M$_\odot$ results in a mass distribution with slope
$-1.28 \pm0.05$. For WB89~380 the slope of a fit to the mass distribution
for $M>6.3$~M$_\odot$ is $-1.49 \pm 0.25$. The slopes derived for these two
outer Galaxy clouds are somewhat on the low side compared to the literature
values quoted above, i.e. the mass distributions are slightly flatter than
what is found in most local clouds.
In part this may be due to the uncertainty introduced by the small number of
clumps in these clouds. If real, this discrepancy might be interpreted as an
indication that (these) outer Galaxy clouds would form relatively more massive
stars than local clouds.
However, fits to the mass distributions of the RMC (high-res.; for
$M>7.9$~M$_\odot$), RMC (low-res.;for $M>126$~M$_\odot$) and
Orion~B South (for $M>2.0$~M$_\odot$) yield slopes of $-1.55 \pm 0.03$, 
$-1.24 \pm 0.09$ and $-1.14 \pm 0.05$, respectively, the latter two of which 
are also flatter than the values found in the literature. In particular, 
Kramer et al. (\cite{kramer}), found a slope of $-1.72\pm 0.09$ for Orion, 
using the same dataset, but a different clump-finding routine GAUSSCLUMPS, 
though we note that many clumps in their list are found lying on and in part 
beyond the edge of the mapped region, or appear to be unresolved, and 
therefore ought to have been excluded from their fit. 
A similar lowish value is found for Maddalena's Cloud (G216$-$2.5; $-$1.4) by 
Williams et al. (\cite{williams}), using the same clump finding method as we 
do here, which suggests this may be due to the properties of CLUMPFIND. 
Compared to GAUSSCLUMPS, and compared to derivations of mass spectra of 
ensembles of clouds (rather than of clumps within a cloud), CLUMPFIND tends to 
underestimate the number of low-mass clumps, lumping them with the more 
massive ones, thus giving rise to shallower mass spectra (Kramer et 
al.~\cite{kramer}). The fact that we have used only resolved clumps in the 
derivation of the mass spectra will strenghten this tendency, because the 
unresolved clumps will also be the ones with the lowest mass.
On the other hand, GAUSSCLUMPS finds small clumps that may not be real, but
parts of larger non-Gaussian clumps.
\noindent
However, neither in the outer Galaxy clouds, 
nor in the local clouds, is the slope of the mass distribution the same as that
of the IMF (i.e. $-$1.5 for stars of mass $1-10$~M$_{\odot}$ and $-$2.3
for stars of mass $>10$~M$_{\odot}$; Miller \& Scalo~\cite{miller}).
For outer Galaxy clouds this situation is even more critical in view of those 
studies (Garmany et al.~\cite{garmany}; Wouterloot et al.~\cite{wouterfieg})
that find that at $R > R_0$ the slope of the IMF becomes steeper, resulting in 
an even larger difference with the values found here.
One can think of several ways in which the observed clump mass spectrum can
evolve into the IMF: The process of star formation is always accompanied
by phases of mass loss; if the mass that is being shed increases with the
mass of the star being formed, this will steepen the slope of the
mass distribution. Alternatively (or in addition), the clumps may contain 
smaller sub-clumps (that are so far unresolved), the mass distribution of 
which may approach that of the IMF; and/or the clumps we have detected will 
undergo further fragmentation before star formation sets in. 

\begin{figure}
\resizebox{\hsize}{!}{\includegraphics{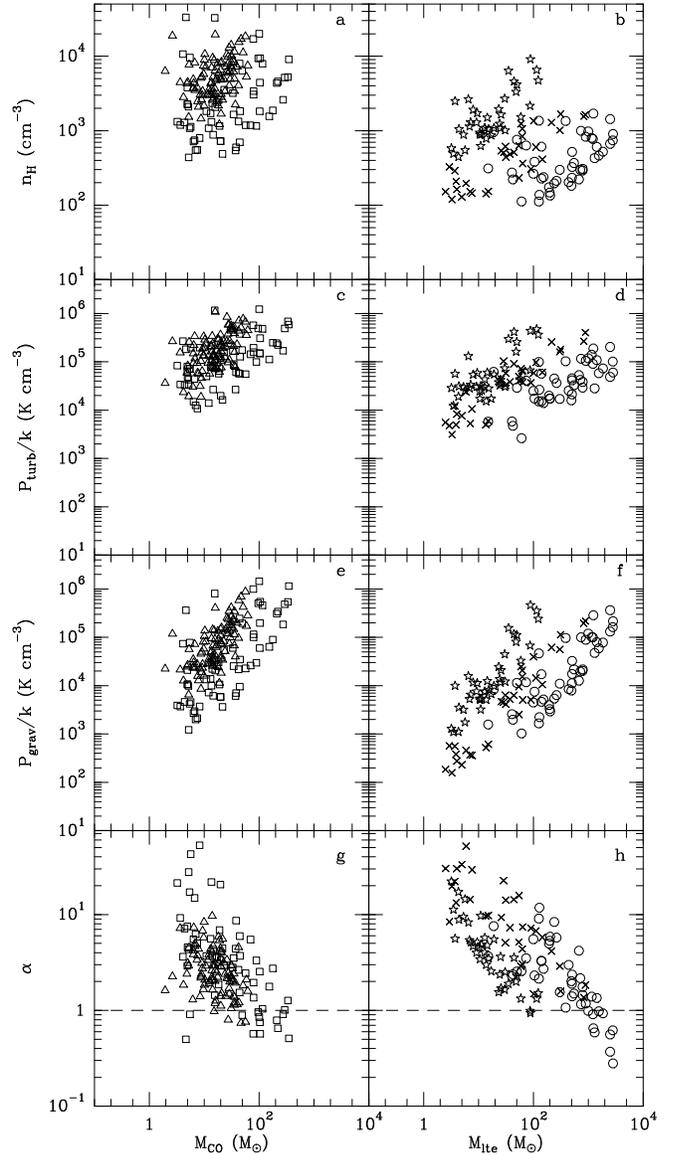}}
\hfill
\caption{Physical parameters of the CO clumps as a function of mass for
two outer Galaxy clouds (left) and two local clouds (right). Symbols are for
WB89~85 (open squares), WB89~380 (open triangles), Orion B South
(crosses), and the RMC low-resolution (open circles) and high-resolution
data (stars).
The following parameters are shown:
{\bf a, b}\ The mean clump gas density $\bar n_{\rm H}$;
{\bf c, d}\ The turbulent pressure $P_{\rm turb}$/k;
{\bf e, f}\ The pressure due to the self-gravity of the clump,
$P_{\rm grav}$/k;
{\bf g, h}\ The virial parameter $\alpha$ (=$P_{\rm turb}/P_{\rm grav}$).}
\label{2by4}
\end{figure}

\subsubsection{Physical parameters \label{physpars}}

We shall compare several parameters of the clumps in the outer Galaxy
clouds with those in the local comparison sample.

\noindent
The total kinetic energy of a clump is $E_{\rm kin} = 3/2 M_{\rm CO} \sigma^2$,
where $\sigma$ is the one-dimensional velocity dispersion, which includes
both the thermal and turbulent components. The gravitational energy,
assuming a spherical mass distribution with density $\rho \propto
r^{-2}$ is $E_{\rm grav}$ = -- G $M_{\rm CO}^2/r$. The magnetic energy, due
to a field of strength $B$, is $B^2$/8$\pi$. The virial theorem for a clump,
with external pressure $P$, can then be written as 

$$4 \pi r^3 P = 3 M_{\rm CO} \sigma^2 - {\rm G} M_{\rm CO}^2/r + B^2/8\pi$$

\noindent
(after Fleck~\cite{fleck}). From this we find

$$P/{\rm k} = \bar \rho \sigma^2 /{\rm k} - {\rm G} M_{\rm CO} \bar \rho /3 r 
{\rm k} + B^2 /8  \pi {\rm k}, $$

\noindent
where $\bar \rho = (3/4\pi) M_{\rm CO} r^{-3}$ is the average mass density.
The first term on the right-hand side in this equation is the pressure due to
the motions of the molecules in the clump, $P_{\rm turb}$/k (which we shall
call `turbulent pressure', because the non-thermal contribution to $\sigma$
dominates); the second term is the gravitational pressure $P_{\rm grav}$/k,
and the last term is the magnetic pressure $P_{\rm magn}$/k. These terms can 
be written as follows: 

$$P_{\rm turb}/{\rm k [Kcm^{-3}]} = 29.66 \times \bar n_{\rm H} \, 
\Delta v^2,$$

\noindent
with $\bar n_{\rm H}{\rm [cm^{-3}]} = \bar \rho / \mu_{\rm H} = 7.12 \times
M_{\rm CO}{\rm [M_\odot]}\, r{\rm [pc]}^{-3}$, the average volume density of
nucleons, $\mu_H$ the mass of a hydrogen atom, corrected for a contribution of
He (a factor 1.36), and $\Delta v$ [kms$^{-1}$] the FWHM line width, and 

$$P_{\rm grav}/{\rm k [Kcm^{-3}]} = 0.236 \times M_{\rm CO}{\rm [M_\odot]}\, 
\bar n_{\rm H}{\rm [cm^{-3}]}\, r{\rm [pc]}^{-1}.$$

\noindent
This last term can also be written in terms of the surface density $\Sigma$
(= $M/\pi r^2$): $P_{\rm grav}/{\rm k = (G \pi /k)} \times \Sigma^2$, or

$$P_{\rm grav}/{\rm k [Kcm^{-3}]} = 16.58 \times \Sigma[M_{\odot} 
{\rm pc}^{-2}]^2.$$

\noindent
The ratio of the turbulent and the gravitational pressures is 

$$\alpha = 3 r \sigma^2 / {\rm G} M_{\rm CO} = 126\, r{\rm [pc]}\, 
\Delta v[{\rm kms}^{-1}]^2 /M_{\rm CO}[{\rm M}_\odot],$$

\noindent
which is identical to the ratio between $M_{\rm vir}$ and $M_{\rm CO}$, and is
usually called the virial parameter (see e.g. Bertoldi \&
McKee~\cite{bertoldi}). If a clump is in virial equilibrium with external
surface pressure $P$=0, and in the absence of a magnetic field, then the
turbulent pressure is exactly balanced by the gravitational pressure,
$\alpha = 1$, and $M_{\rm CO}$ = $M_{\rm vir}$. If the kinetic energy is
exactly balanced by its gravitational energy, $E_{\rm kin}$ =
G$M_{\rm CO}^2/r$, then its mass is $M_{\rm CO} = 3 r \sigma^2/2 {\rm G} = 
0.5 \times M_{\rm vir}$; the cloud is then called gravitationally bound.

\noindent
The magnetic pressure can be written as

$$P_{\rm magn}/{\rm k [Kcm^{-3}]} = 288.3 \times B[\mu {\rm G}]^2.$$

\medskip\noindent
In Fig.~\ref{2by4} we have plotted the clump parameters derived
above, as a function of clump mass. In this figure, the panels on the
left-hand side are for the outer Galaxy clouds, while those on the
right-hand side are for the local comparison clouds.
The average densities of the clumps in the local clouds lie more or less
between 10$^2$ and 10$^4$~cm$^{-3}$, with the clumps found in the
high-resolution IRAM 30-m RMC maps having the highest values at a fixed mass;
for the clumps in our outer Galaxy sample, the density ranges between
$4 \times 10^2$ and $4 \times 10^4$~cm$^{-3}$, with those in WB89~380 having
the highest values. WB89~380 has been observed at the highest spatial
resolution, and has more clumps with smaller radii; the density
$\bar n_{\rm H}  \propto M_{\rm CO} \times r^{-3} \propto r^{-1}$, and thus
higher densities can be expected.
For the majority of clumps in the outer Galaxy clouds, the turbulent gas
pressure ($10^4 - 10^6$~Kcm$^{-3}$; panel~c) and the pressure due to
self-gravity ($10^3 - 10^6$~Kcm$^{-3}$; panel~e) are both higher than those
in local clumps at comparable mass.
Also here, for the local clouds at fixed clump mass, the highest pressures
are for the higher-resolution RMC clumps.
The virial parameter, i.e. the ratio of turbulent (trying to disperse the
clumps) and gravitational (working to keep them together) pressure
(panels g, h), also shows a different distribution in both cloud samples. The
most massive clumps in both outer Galaxy- and local clouds have
$\alpha \approx$1, i.e. there is equilibrium between $P_{\rm turb}$ and 
$P_{\rm grav}$, but for the lower-mass clumps the turbulent pressure is the 
larger one, and $\alpha$ increases with decreasing clump mass, implying that 
the lower-mass clumps would need some external pressure if they are to be 
confined.
The important thing to note, however, is that in the outer Galaxy clouds there
are a number of low-mass clumps (down to the lowest masses found; a few
M$_{\odot}$) where self-gravity is the dominating force balancing the internal
turbulent pressure (i.e. $\alpha \approx$1), which suggests that these are 
potentially star-forming clumps. We shall return to this interesting fact 
later.

\subsubsection{Clump stability and implications for the IMF \label{clumpstab}}

After the general considerations presented in Sect.~\ref{physpars}, 
in this section we consider the stability of the clumps, i.e. we look at the
balance between the various pressure terms (those trying to expand the clump, 
and those working to compress it). 

\begin{figure}
\resizebox{\hsize}{!}{\includegraphics{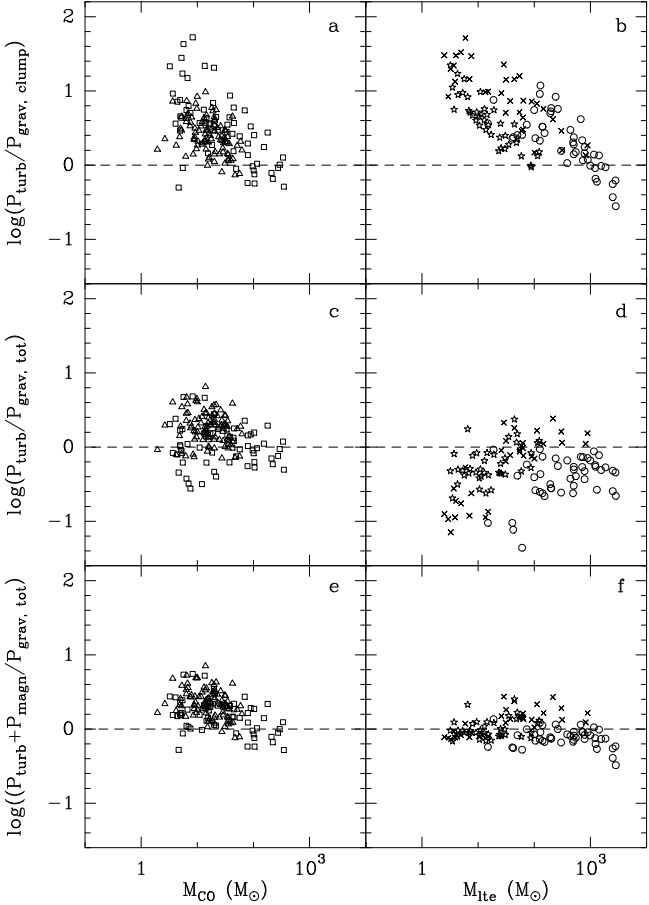}}
\hfill
\caption{Ratio of the pressures balancing a clump's equilibrium for clumps in
outer Galaxy clouds (left-hand panels) and in local clouds (right). Symbols
are as follows: WB89~380, open triangles; WB89~85: open squares. Orion B 
South: crosses; RMC lower-resolution data: open circles; RMC higher-resolution 
data: stars.
{\bf a, b}\ Ratio of turbulent and gravitational pressures; {\bf c, d}\ As
a, b but the gravitational pressure is the sum of the pressure due to the
clump's self-gravity, and the pressure on the clumps exerted by the cloud's
interclump gas, as a consequence of the self-gravity of the cloud as a
whole; {\bf e, f}\ As c, d, but the pressure term of a 10$\mu$G magnetic
field has been added to the turbulent pressure.}
\label{prats}
\end{figure}

\smallskip\noindent
A molecular cloud is an ensemble of relatively high-density clumps, which are 
moving about in a low-density (interclump) medium. Both the movement of the
clumps and the gravitational force of the molecular cloud material exert
pressures on the interclump medium. If the cloud as a whole is in
equilibrium, these pressures are of comparable magnitude and work in
opposing directions. The clumps themselves are also subject to various
pressures. The thermal and (the larger) turbulent motions of the molecular 
material within them seek to expand the clumps, while their self-gravity
and the pressure of the surrounding interclump medium work against this. 
The clumps are likely to be threaded by a magnetic field, which also opposes 
compression (at least perpendicular to the field lines).

\smallskip\noindent
$P_{\rm turb}$ and $P_{\rm grav}$ for the individual clumps were calculated and
shown in Sect.~\ref{physpars} (Fig.~\ref{2by4}).
The interclump pressure may be calculated by assuming that it is in
equilibrium with the gravitational pressure exerted by the cloud as a whole.
This is in some sense dependent on the pressure of the
general ISM surrounding the clouds, which is about 10--20 times lower at
$R \approx 17$~kpc as it is in the solar neighbourhood (BW95). This lower
pressure allows outer Galaxy clouds to relax to a larger radius than clouds of
the same mass at smaller $R$ (BW95), and because the gravitational pressure of
a cloud depends also on its radius, the general ISM pressure influences the
pressure of the interclump gas. For the clouds under consideration here, this
pressure is of the order of several $10^4$~Kcm$^{-3}$:
$P_{\rm grav, cloud}$/k $\approx 3.7\times 10^4$~Kcm$^{-3}$ (WB89~85), 
$1.7\times 10^4$~Kcm$^{-3}$ (WB89~380), $4.4\times 10^4$~Kcm$^{-3}$
(Orion), and $5.9\times 10^4$~Kcm$^{-3}$ (RMC). For the outer Galaxy clouds
we considered the mass and radius of the region in which the BIMA observations
were carried out, rather than the mass and radius of the entire cloud as
detected with KOSMA (a similar procedure has been applied to the Orion and
Rosette clouds).

\noindent
Typical values of the magnetic field in molecular clouds are of the order of
a few 10$\mu$G (Troland \& Heiles~\cite{troland}), which leads to
$P_{\rm magn}$/k $\approx 2.9\times 10^4$~Kcm$^{-3}$.

\medskip\noindent
In Fig.~\ref{prats} we look at the combination of these pressures that
influence the stability of a clump. Here we first show (Fig.~\ref{prats}a, b)
the ratio of the clumps' turbulent pressure and the pressure due to their
self-gravity. The situation is the same as what is shown in Fig.~\ref{2by4}g,
h and has been described in Sect.~\ref{physpars}. In Fig.~\ref{prats}c, d we
plot the ratio of the former and the {\it total} gravitational pressure, i.e.
including that which is exerted by the cloud as a whole, through the
interclump medium, which shows the consequences on the pressure balance of
including an external pressure term. Figs.~\ref{prats}c, d demonstrate that
when the pressure of the interclump medium is also taken into account,
virtually all clumps are in or close to equilibrium, i.e. the clumps' own
gravity and that of the cloud as a whole, are balanced by the clumps'
turbulent pressure. In fact, for many clumps in the local sample
the interclump pressure is much larger than the turbulent pressure. Taken at
face value this would suggest large-scale clump collapse and star formation,
which is clearly not what we observe, and to prevent large-scale collapse a
stabilizing agent is clearly needed. It is also clear that the influence of
the interclump pressure on outer Galaxy clumps is smaller than on the local
clumps. This is because the interclump pressure is of the same order in both
outer and inner Galaxy clouds, but $P_{\rm turb}$ is larger in the outer 
Galaxy clumps.

\noindent
Figs.~\ref{prats}e, f show the consequence of adding the pressure term due to 
a 10$\mu$G magnetic field: for nearly all clumps equilibrium has been 
restored, with respect to the situation shown in Figs.~\ref{prats}c, d. It may 
be that the actual value of the magnetic field is larger, especially in the 
more massive clumps and in the local clouds, but already with the presently 
used low value it is clear that the magnetic field is an important ingredient 
in the pressure balance. 

\smallskip\noindent
From what we have found in Fig.~\ref{prats}a, b, and from the behaviour
of the virial parameter $\alpha$ in Figs.~\ref{2by4}g, h, it appears
that in these outer Galaxy clouds gravity may be the dominant force down to a
lower mass than in local clouds. In fact, a closer look at
Figs.~\ref{2by4}g, h reveals that in the IRAM-RMC data the lowest-mass clump
with $\alpha \leq 2$ has $M \approx23.4$~M$_{\odot}$, while this is
2.0~M$_{\odot}$ and 4.7~M$_{\odot}$ in WB89~380 and WB89~85, respectively,
i.e. up to an order of magnitude smaller. Moreover, in WB89~380 there are 14
clumps with $M<23.4$~M$_{\odot}$ and $\alpha \leq 2$; in WB89~85 there are 5
clumps that fall within these constraints. This implies that
gravitational collapse and star formation may occur more readily even in the
smallest of these outer Galaxy clumps. Since a clump of mass $M$ can only form
stars of mass less than $M$, this would mean that an excess of low mass 
stars is expected to form in the outer Galaxy with respect to local clouds. 
This would provide an explanation for the results of Garmany et al.
(\cite{garmany}), and Wouterloot et al. (\cite{wouterfieg}) who found that 
the IMF steepens in the outer Galaxy. 

\smallskip\noindent
We note that in the above, notably in Figs.~\ref{sizlinall}a, and \ref{2by4},
there is a separation between the data for WB89~380 and WB89~85: at the same
$\Delta v$, the majority of WB89~380 clumps have a smaller radius than the
WB89~85 clumps, while at identical $M_{\rm CO}$ they have a higher density,
and a higher internal pressure.
A naive interpretation is that this illustrates the difference in evolutionary 
stage between the two clouds: the younger one, WB89~380, has mostly smaller,
higher density clumps, that are on the brink of gravitational equilibrium,
and are about to form stars, while the older region WB89~85 has been left
with lower-density, more stable clumps (except perhaps for the most massive
ones). However, there is a resolution effect to consider: the WB89~85 cloud
was observed with a beam size of $0.60\times 0.47$~pc$^2$, and a map was
constructed on a grid with 0.11~pc spacing; CLUMPFIND looked for clumps with 
a resolution of 0.22~pc in both coordinates (2$\times$2 grid resolution 
elements); WB89~380 was observed with a beam of $0.36\times 0.23$~pc$^2$, 
a map was constructed on a grid with 0.11~pc spacing, and clumps were searched 
with a resolution of 0.14$\times$0.07~pc$^2$ (2$\times$1 grid resolution 
elements), and thus CLUMPFIND can come up with smaller clumps in WB89~380. 
The same effect is found in the local clouds, from the Williams et al. 
(\cite{willblitz}) RMC observations, via Orion, to the Schneider et al. 
(\cite{schneider}) RMC observations, there is an increase in resolution, and a 
corresponding shift in the distributions of the data points towards smaller 
radii and masses, and (at fixed mass) towards higher densities and pressures. 
The grid-resolution of the Schneider et al. data (0.12~pc) is comparable to 
that of WB89~85 and WB89~380, but the beam size of the Schneider et al. 
observations (0.09~pc) is smaller, though close to that of the WB89~380 
observations.

\noindent
The question is now, whether the difference we see in Figs~\ref{2by4}g, h
between outer Galaxy and local (especially the IRAM-RMC) clumps is also a
consequence of resolution. Resolution will certainly play a role, as can be
seen from the difference in distributions for the RMC high- and
low-resolution data, and Orion. Whether resolution is the {\it only}
explanation is more difficult to say, because this would imply having to
demonstrate that the type of clumps (of this size and mass) we see in the
outer Galaxy clouds (and then especially in WB89~380) could also be seen in
the high-resolution RMC data. We consider the following: in spite of the
IRAM beam being smaller than (with respect to the WB89~85 observations) or
comparable to (with respect to the WB89~380 observations) those used in the
BIMA observations, we do find clumps of comparable mass at the low end of
the mass spectrum. The smallest clump mass found in the IRAM-RMC data has
$M\approx3.2$~M$_{\odot}$ (cf. 3.3~M$_{\odot}$ in WB89~85, and
2.0~M$_{\odot}$ in WB89~380). A plot of radius versus mass (not shown) shows
that at fixed mass, the radii of the IRAM-RMC clumps is larger than those of
WB89~380 clumps (in spite of the IRAM beam being smaller than the BIMA beam),
and similar to those of WB89~85 clumps, while the linewidths are comparable.
Only at $M<10$~M$_{\odot}$ the linewidths of the WB89~380 clumps are smaller. 
As a consequence, at fixed mass, $\alpha = M_{\rm vir}/M_{\rm CO}$ is smaller 
for the WB89~380 clumps than for the IRAM-RMC clumps, giving rise to the 
effect seen in Figs.~\ref{2by4}g, h. 

\noindent
Masses for the outer Galaxy clumps have been calculated using the empirical
method of $M_{\rm CO} = X \times \int T{\rm d}v$, where for $X$ we have used
the local value of 1.9 $\times 10^{20}$~(Kkms$^{-1}$)$^{-1}$cm$^{-2}$. In BW95
we argued that even in the far-outer Galaxy this results in masses equal to
LTE masses (derived from $^{13}$CO observations) if the latter are calculated
using the [$^{13}$CO]/[H$_2$] abundance appropriate for that region of the
Galaxy. This is in contrast to claims found in the literature, which would
have $X$ at large distances from the galactic center $\sim$2--4 times larger
than the local value (e.g. Digel et al.~\cite{digel}; May et al.~\cite{may}).
If for the sake of argument we would adopt an $X$-value of three times the
local value, while $r$ and $\Delta v$ remain the same (these are 
$X$-independent of course), we find that virually all clumps in WB89~380 have 
$\alpha \leq 2$, while the lowest clump mass is still only 6~M$_{\odot}$, i.e. 
almost 4 times smaller than the lowest clump mass in the IRAM-RMC observations 
having $\alpha \leq 2$, and thus still resulting in the same conclusion, 
namely that in (these) outer Galaxy clouds gravity is the dominant force down 
to a lower mass than in (these) local clouds.

\smallskip\noindent
While for the outer Galaxy clumps we arrive at equivalent LTE masses through
the use of $X$, LTE masses for the local clumps were derived directly from the
column densities of $^{13}$CO, and are not expected to show a {\it
systematic} difference with respect to the masses of the outer Galaxy clumps.
A potentially serious caveat is whether $^{12}$CO observations
allow one to accurately identify clumps (compared to $^{13}$CO), i.e. density
rather than temperature enhancements. An indication of this is perhaps given
by the fact that the CO and CS clumps in general do not coincide. This may
however be because in CS we might not be seeing clumps at all, but rather
different parts of the expanding shell (see Sect.~\ref{results}), that show up
at different velocities, hence giving the impression of clumpy structure where
none may exist. It may also be due to the different excitation conditions of
these molecules. In fact, from a (low-resolution) CO survey of GMCs
near Cas~A and NGC~7538, Ungerechts et al. (\cite{ungerechts}) find that the
distribution of $^{12}$CO and $^{13}$CO is very similar, and that the
$\int T{\rm d}v$ of both molecules are closely correlated at each grid point
in the map. Radii derived from $^{13}$CO observations are usually smaller,
however, as are the line widths (though not necessarily so in a turbulent
medium; see Ungerechts et al.~\cite{ungerechts}, who also find that
$M_{\rm vir}$ from $^{12}$CO and $^{13}$CO agree within a factor of two),
affecting most parameters derived here.
Whether $^{13}$CO-derived radii are smaller than those derived from
$^{12}$CO, also depends on the relative sensitivity of the observations.
In the inner Galaxy, the antenna temperature ratio of $^{12}$CO/$^{13}$CO
$\approx$3 (Gordon \& Burton~\cite{gordon}); the (2--1)/(1--0) line ratios
are approximately unity. The 2.5$\sigma \sim 1.3$~K limit used in the clumpfind
procedure for the high-resolution RMC $^{13}$CO data would correspond to
$T(^{12}$CO) of about 4~K. According to BW95 there is no gradient in the mean
cloud antenna temperature with distance from the galactic center, and it
should be possible to compare this temperature directly with that which we
used as limiting value for the clumpfind procedure in the outer Galaxy clouds,
viz. 3.3~K for WB89~85 and 8.7~K for WB89~380. From this we see that WB89~85
and the IRAM-observations of the RMC are compatible, while we go less deep in
the case of WB89~380. This might result in seeing only the `tips of the
icebergs' in WB89~380, i.e. finding only the peaks of the clumps, resulting
in fewer clumps, which are also smaller and less massive than they would be,
had we been able to search down to a lower level. To check how this
influences our results, we ran a clumpfind-test on the WB89~85 data, using the
same limiting value as we have for WB89~380, i.e. 8.7~K. As expected,
CLUMPFIND now finds fewer clumps (34, 12 of which are considered useful
[resolved and not at the edge of the map]).
The new clumps are found at the positions of the most massive clumps from our
original analysis of this cloud, with masses, radii, and line widths for
most clumps larger than the clumps originally found at these positions.
In a mass-radius plot the old and new data points overlap.
The clump with the lowest mass has $M_{\rm CO} \sim 19$~M$_{\odot}$, and 
$\alpha \sim 3$. The others have masses between 84.5 and 586~M$_{\odot}$, and
$\alpha$ between 0.7 and 2.1. Because we do not find smaller and less
massive clumps by going less deep, this result strengthens the conclusion 
that the difference found between the high-resolution RMC results and those of 
the outer Galaxy clouds is not due to the fact that different transitions were
observed, nor due to sensitivity.

\noindent
In spite of the arguments presented above, it is clear that comparing such
different data sets is not a straight-forward matter, because many effects
play a role. One would need to study a sample of local
and outer Galaxy clouds with carefully selected telescopes, in the same
transitions and analyzed in the same way. The cloud sample should preferably
be selected taking into account $L_{\rm fir}$ of the IRAS source, and
evolutionary status, to ensure that as little bias as possible is introduced.

\section{Summary \label{summ}}

The fields around luminous IRAS point sources in three molecular clouds in the
outer Galaxy have been observed in CO (1--0) and CS (2--1) using the BIMA
interferometer. These observations were combined with single dish data from 
the NRAO 12-m and the IRAM 30-m telescopes. Using this technique we were able 
to obtain maps around these sources, of about 6 pc in diameter with a 
resolution of about 0.2 pc. This enables us to analyse 
small-scale properties of the molecular gas on scales similar to the smallest
size-scales on which clump properties of molecular clouds in the solar
neighbourhood are studied.  The data sets have been analyzed using an automated
three-dimensional clump-finding and -analysis program. 

\noindent
From the CO data we find a size-linewidth relation that is steeper than what
is found in nearby molecular clouds. This can be explained by the rather
narrow range in radius for the clumps in our data set (Sect.~\ref{sizlinco}).
From the CO data we find a somewhat shallower clump mass distributions than in
nearby clouds, with slopes of $-1.28 \pm 0.05$ (WB89~85) and $-1.49 \pm 0.25$
(WB89~380) in plots of log(d$N$/d$M$)\ vs.\ log$M$. However, these slopes 
are consistent with those found by Williams et al. (\cite{williams}) for the 
RMC and G216$-$2.5 (Maddalena's Cloud), derived from clumps identified with the
same method as used here, and is possibly due to the fact that this method
tends to underestimate the number of low-mass clumps, resulting in shallower
mass distributions (Sect.~\ref{massdist}).

\smallskip\noindent
In Sect.~\ref{physpars} and \ref{clumpstab} we look in detail at the
equilibrium of the clumps both in the two of the outer Galaxy clouds
(WB89~85 and WB89~380) and in two local comparison clouds (RMC and Orion B
South). The most notable finding is that for outer Galaxy clumps down to the
lowest (a few M$_{\odot}$) masses found turbulent and gravitational pressure
are in equilibrium (virial parameter $\alpha \approx 1$). For clumps in the
local clouds this is only so for the more massive (a few tens of
M$_{\odot}$) ones. Adding the pressure of the interclump medium and that,
due to the magnetic field (using a representative value of 10~$\mu$G) to the
equation, brings virtually all clumps in the clouds under consideration into
equilibrium. We note that the interclump pressure in the outer Galaxy clouds
has relatively less influence on the balance, compared to the local clouds,
because in the outer Galaxy clouds $P_{\rm turb}$ of the clumps is several 
times larger than in the local clumps, while the interclump pressure is of the
same magnitude. 

\noindent
These results indicate that in these outer Galaxy clouds gravity may be the
dominant force down to a much lower mass than in local clouds, implying that
gravitational collapse and star formation may occur more readily even in the
smallest outer Galaxy clumps. As this leads to the formation of relatively
more low-mass stars, it would provide an explanation for the observed
steepening of the IMF in the outer Galaxy.

\noindent
The results presented above are necessarily based on an inhomogeneous
dataset, and a comparison between local and outer Galaxy clouds ought to be
repeated with a more carefully selected sample using a uniform method of
analysis. The potential importance of the conclusion based on the present
data makes such a renewed study worthwhile.

\begin{acknowledgements} 
EdG, JB, and JGAW acknowledge support from NATO grant 920835. EdG acknowledges
financial support from NSF grant AST-8918912 and NASA grant NAG 5-1736. ALR
acknowledges the support of the NSF Young Faculty Career Development CAREER
Program via NSF grant 96-24924.
This work was completed while JGAW was a Visiting Professor at the Istituto
di Radioastronomia, CNR, Bologna. We thank Jonathan Williams, Carsten
Kramer, and Nicola Schneider for generously making their original data 
available for
analysis, and Malcolm Walmsley and Leo Blitz for commenting on an earlier 
version of this paper. Jonathan Williams is also thanked for providing us
with the latest version of CLUMPFIND, running in IDL, and for patiently
answering our questions about this procedure.
\hfill\break\noindent
The KOSMA radio telescope at Gornergrat-S\"ud Observatory is operated by the
University of K\"oln, and supported by the Deutsche Forschungsgemeinschaft
through grant SFB-301, as well as by special funding from the Land
Nordrhein-Westfalen. The Observatory is administered by the Internationale
Stiftung Hochalpine Forschungsstationen Jungfraujoch und Gornergrat, Bern,
Switzerland.
\end{acknowledgements}

\begin{table*}
\caption[]{CO Clump parameters of WB89~85.
Right ascension and declination are in
arcsecond offsets from the central position of the map: 
$(\alpha,\delta)_{1950}$ = (21$^h$ 27$^m$ 5\fs9 , +54\degr\ 23\arcmin\ 42\arcsec).
\label{s127cocl}}
\begin{flushleft}
\begin{tabular}{rrrcrrcccr}
\hline\noalign{\smallskip}
\multicolumn{1}{c}{Clump}& \multicolumn{1}{c}{$\Delta$RA}& 
\multicolumn{1}{c}{$\Delta$Dec}&
\multicolumn{1}{c}{$V_{\rm lsr}$} & \multicolumn{1}{c}{$T_{\rm peak}$} &
$T_{\rm ave}$ & $r_{\rm eff}$ & $\Delta v_{\rm fwhm}$ & $M_{\rm CO}$ &
$M_{\rm vir}$ \\
\multicolumn{1}{l}{}& \multicolumn{1}{c}{[\arcsec]}&
\multicolumn{1}{c}{[\arcsec]}& [km$\,$s$^{-1}$] & \multicolumn{1}{c}{[K]} &
\multicolumn{1}{c}{[K]} & [pc] & [km$\,$s$^{-1}$] & 
[M$_\odot$] & [M$_\odot$]\\
\multicolumn{1}{c}{(1)} & \multicolumn{1}{c}{(2)} & \multicolumn{1}{c}{(3)} &
\multicolumn{1}{c}{(4)} & \multicolumn{1}{c}{(5)} & \multicolumn{1}{c}{(6)} &
\multicolumn{1}{c}{(7)} & \multicolumn{1}{c}{(8)} & \multicolumn{1}{c}{(9)} &
\multicolumn{1}{c}{(10)} \\
\hline\noalign{\smallskip}
   1 & 24. & 50. &-92.50 &57.3 &23.2 &0.65 &1.48 &349.0 &179.0\\
   2 & 28. & 52. &-93.00 &54.7 &25.7 &0.43 &1.05 &105.0 & 59.7\\
   3 & 30. & 52. &-93.50 &44.4 &17.7 &0.70 &1.28 &221.0 &145.0\\
   4 &  8. & 36. &-93.00 &40.0 &15.3 &0.77 &2.10 &336.0 &428.0\\
   5 & 58. &-24. &-94.00 &39.6 &16.8 &0.42 &1.33 & 96.8 & 93.6\\
   6 & 58. &-30. &-94.50 &37.1 &20.3 &0.33 &1.44 &101.0 & 86.2\\
   7 & 52. &-32. &-95.75 &36.0 &17.3 &0.55 &1.06 &104.0 & 77.9\\
   8 & 54. &-38. &-95.00 &35.1 &17.2 &0.47 &1.43 &116.0 &121.0\\
   9 &-22. &-52. &-93.50 &34.4 &15.8 &0.70 &1.37 &210.0 &166.0\\
  10 & 12. & 40. &-94.00 &33.0 &13.8 &0.91 &1.48 &275.0 &251.0\\
  11 &-20. &-32. &-93.75 &32.9 &13.9 &0.74 &1.79 &294.0 &299.0\\
  12 & 60. &-42. &-96.50 &29.0 &15.8 &0.32 &1.06 & 78.7 & 45.3\\
  13 & 48. &-24. &-96.25 &25.4 & 9.4 &0.89 &1.56 &154.0 &273.0\\
  14 &  4. & -4. &-92.50 &23.7 & 7.8 &0.88 &2.11 &179.0 &494.0\\
  15 & 54. &-30. &-96.75 &21.8 &11.2 &0.38 &0.96 & 49.3 & 44.1\\
  16 &-58. &-48. &-93.50 &20.5 & 8.8 &0.34 &1.26 & 44.2 & 68.0\\
  17 & -4. &-54. &-95.50 &20.0 & 9.4 &0.55 &1.25 & 79.6 &108.0\\
  18 & 36. & 56. &-91.25 &17.2 & 9.8 &0.42 &1.33 & 33.3 & 93.6\\
  19 &-42. &-44. &-92.75 &16.5 & 9.0 &0.57 &1.47 & 79.4 &155.0\\
  20 & -6. & 44. &-92.75 &16.3 & 7.3 &0.74 &1.68 &103.0 &263.0\\
  21 & 58. &-46. &-91.50 &16.0 & 7.5 &0.15 &1.09 & 15.6 & 22.5\\
  22 & 28. & 44. &-95.75 &15.8 & 7.3 &0.70 &1.26 & 57.0 &140.0\\
  23 &-42. & 58. &-88.25 &15.0 & 6.9 &0.10 &0.43 &  4.7 &  2.3\\
  24 & 50. &-50. &-94.25 &15.0 & 6.9 &0.56 &1.54 & 45.1 &167.0\\
  25 &-48. &-56. &-88.25 &14.4 & 6.4 &0.16 &0.50 &  5.5 &  5.0\\
  26 &-32. &-50. &-92.50 &14.4 & 7.1 &0.53 &1.35 & 34.5 &122.0\\
  27 & 54. &-50. &-92.00 &14.3 & 6.9 &0.36 &1.23 & 23.2 & 68.6\\
  28 & 56. &-40. &-97.75 &13.8 & 8.2 &0.37 &0.86 & 18.4 & 34.5\\
  29 & 38. & 48. &-90.75 &13.4 & 7.0 &0.37 &1.16 & 17.1 & 62.7\\
  30 &-10. & 14. &-94.25 &13.1 & 6.8 &0.82 &1.70 & 89.6 &299.0\\
  31 &-32. & 26. &-91.50 &12.9 & 6.3 &0.61 &1.21 & 42.4 &113.0\\
  32 &-36. &-14. &-92.75 &12.3 & 5.8 &0.75 &2.02 & 70.1 &386.0\\
  33 & 16. &-56. &-95.50 &12.1 & 5.5 &0.35 &1.09 & 14.1 & 52.4\\
  34 &-46. & 22. &-96.00 &11.2 & 5.2 &0.21 &0.74 &  4.9 & 14.5\\
  35 &-58. & 46. &-89.25 &11.1 & 5.9 &0.14 &0.92 &  4.1 & 14.9\\
  36 &-12. & 44. &-91.25 &11.0 & 5.6 &0.42 &2.37 & 13.6 &297.0\\
  37 &-46. & 22. &-93.50 &10.9 & 5.3 &0.43 &0.66 & 12.1 & 23.6\\
  38 &-14. & 12. &-95.00 &10.8 & 5.4 &0.35 &0.96 & 11.0 & 40.6\\
  39 &-42. & -6. &-92.25 &10.7 & 5.1 &0.49 &1.46 & 19.1 &132.0\\
  40 & -2. &-58. &-97.00 &10.7 & 5.1 &0.26 &0.83 &  5.2 & 22.6\\
  41 &-16. &-28. &-95.50 &10.5 & 5.6 &0.77 &1.66 & 44.7 &267.0\\
  42 &-40. & 46. &-88.50 &10.4 & 5.0 &0.40 &0.84 &  6.4 & 35.6\\
  43 & 12. &-52. &-97.75 &10.3 & 5.2 &0.25 &0.85 &  5.0 & 22.8\\
  44 & -6. & 42. &-91.50 &10.2 & 6.1 &0.58 &0.96 & 20.0 & 67.4\\
  45 &-54. &-10. &-95.75 & 9.8 & 5.8 &0.38 &1.09 & 16.0 & 56.9\\
  46 & 54. & 16. &-97.25 & 9.5 & 5.5 &0.68 &1.06 & 21.6 & 96.3\\
  47 &-36. & 44. &-90.25 & 8.9 & 4.9 &0.46 &0.81 &  7.6 & 38.0\\
  48 &-40. & 20. &-94.75 & 8.9 & 4.6 &0.49 &1.01 & 14.6 & 63.0\\
  49 &  4. &-24. &-93.25 & 8.3 & 4.5 &0.58 &2.36 & 19.8 &407.0\\
  50 &-26. & 18. &-95.00 & 8.1 & 4.2 &0.33 &1.50 &  5.4 & 93.6\\
\noalign{\smallskip}
\hline
\end{tabular}
\end{flushleft}
\end{table*}

\addtocounter{table}{-1}

\clearpage
\begin{table*}
\caption[]{({\it continued})
}
\begin{flushleft}
\begin{tabular}{rrrcrrcccr}
\hline\noalign{\smallskip}
\multicolumn{1}{c}{Clump}& \multicolumn{1}{c}{$\Delta$RA}& 
\multicolumn{1}{c}{$\Delta$Dec}&
\multicolumn{1}{c}{$V_{\rm lsr}$} & \multicolumn{1}{c}{$T_{\rm peak}$} &
$T_{\rm ave}$ & $r_{\rm eff}$ & $\Delta v_{\rm fwhm}$ & $M_{\rm CO}$ &
$M_{\rm vir}$ \\
\multicolumn{1}{l}{}& \multicolumn{1}{c}{[\arcsec]}&
\multicolumn{1}{c}{[\arcsec]}& [km$\,$s$^{-1}$] & \multicolumn{1}{c}{[K]} &
\multicolumn{1}{c}{[K]} & [pc] & [km$\,$s$^{-1}$] & 
[M$_\odot$] & [M$_\odot$]\\
\multicolumn{1}{c}{(1)} & \multicolumn{1}{c}{(2)} & \multicolumn{1}{c}{(3)} &
\multicolumn{1}{c}{(4)} & \multicolumn{1}{c}{(5)} & \multicolumn{1}{c}{(6)} &
\multicolumn{1}{c}{(7)} & \multicolumn{1}{c}{(8)} & \multicolumn{1}{c}{(9)} &
\multicolumn{1}{c}{(10)} \\
\hline\noalign{\smallskip}
  51 &-12. & 16. &-91.50 & 8.0 & 4.8 &0.79 &1.29 & 37.5 &166.0\\
  52 &-44. & 14. &-97.00 & 7.9 & 4.6 &0.26 &1.46 &  3.3 & 69.8\\
  53 &  6. &-60. &-98.25 & 7.9 & 4.7 &0.29 &0.93 &  4.4 & 31.6\\
  54 & 52. & 26. &-98.00 & 7.9 & 4.7 &0.25 &1.09 &  4.9 & 37.4\\
  55 & 22. & 36. &-98.25 & 7.4 & 4.3 &0.44 &1.62 &  5.2 &145.0\\
  56 &-50. & -2. &-91.25 & 7.4 & 4.6 &0.28 &0.98 &  3.7 & 33.9\\
  57 &-40. & 12. &-95.75 & 7.3 & 4.3 &0.40 &1.41 &  6.7 &100.0\\
  58 &-18. & -2. &-95.75 & 7.3 & 4.2 &0.33 &2.42 &  5.7 &244.0\\
  59 &-18. & -8. &-96.00 & 6.9 & 4.3 &0.42 &2.88 &  8.3 &439.0\\
  60 &-26. & 34. &-92.75 & 6.7 & 4.3 &0.45 &0.88 &  7.0 & 43.9\\
  61 &-44. & 22. &-92.50 & 6.7 & 4.9 &0.38 &1.24 & 13.1 & 73.6\\
  62 & 18. &  6. &-90.75 & 6.6 & 4.6 &0.77 &1.85 & 38.3 &332.0\\
\noalign{\smallskip}
\hline
\end{tabular}
\end{flushleft}
\end{table*}

\clearpage
\begin{table*}
\caption[]{CO Clump parameters of WB89~380.
Right ascension and declination are in
arcsecond offsets from the central position of the map: 
$(\alpha,\delta)_{1950}$ = (1$^h$ 4$^m$ 35\fs7 , +65\degr\ 5\arcmin\ 21\arcsec).
\label{380cocl}}
\begin{flushleft}
\begin{tabular}{rrrcrrcccr}
\hline\noalign{\smallskip}
\multicolumn{1}{c}{Clump}& \multicolumn{1}{c}{$\Delta$RA}& 
\multicolumn{1}{c}{$\Delta$Dec}&
\multicolumn{1}{c}{$V_{\rm lsr}$} & \multicolumn{1}{c}{$T_{\rm peak}$} &
$T_{\rm ave}$ & $r_{\rm eff}$ & $\Delta v_{\rm fwhm}$ & $M_{\rm CO}$ &
$M_{\rm vir}$ \\
\multicolumn{1}{l}{}& \multicolumn{1}{c}{[\arcsec]}&
\multicolumn{1}{c}{[\arcsec]} & [km$\,$s$^{-1}$] & \multicolumn{1}{c}{[K]} &
\multicolumn{1}{c}{[K]} & [pc] & [km$\,$s$^{-1}$] & 
[M$_\odot$] & [M$_\odot$]\\
\multicolumn{1}{c}{(1)} & \multicolumn{1}{c}{(2)} & \multicolumn{1}{c}{(3)} &
\multicolumn{1}{c}{(4)} & \multicolumn{1}{c}{(5)} & \multicolumn{1}{c}{(6)} &
\multicolumn{1}{c}{(7)} & \multicolumn{1}{c}{(8)} & \multicolumn{1}{c}{(9)} &
\multicolumn{1}{c}{(10)} \\
\hline\noalign{\smallskip}
   1 & 38. & 18. &-87.00 &49.5 &24.5 &0.28 &1.11 & 56.7 & 43.5\\
   2 & 41. &  3. &-87.00 &45.1 &20.4 &0.27 &1.24 & 43.0 & 52.3\\
   3 & 39. &  6. &-87.25 &44.7 &26.8 &0.27 &1.01 & 47.0 & 34.7\\
   4 & 42. &-38. &-88.25 &41.9 &19.4 &0.34 &1.07 & 34.4 & 49.0\\
   5 & 39. & -3. &-87.50 &41.1 &21.7 &0.25 &1.02 & 31.1 & 32.8\\
   6 & 42. &-15. &-87.50 &40.3 &18.4 &0.34 &1.04 & 39.8 & 46.3\\
   7 &  6. & -3. &-88.50 &39.6 &17.3 &0.39 &1.42 & 62.4 & 99.1\\
   8 & 29. & 42. &-87.50 &38.9 &13.9 &0.28 &0.90 & 13.7 & 28.6\\
   9 & 39. &  0. &-88.00 &38.8 &19.4 &0.32 &0.74 & 27.4 & 22.1\\
  10 & 42. &-11. &-87.00 &38.3 &16.4 &0.25 &1.24 & 28.5 & 48.4\\
  11 & 42. &-33. &-87.25 &37.9 &19.7 &0.37 &1.03 & 35.7 & 49.5\\
  12 & 29. & 24. &-87.25 &37.5 &18.1 &0.32 &1.53 & 51.4 & 94.4\\
  13 &  6. & -2. &-87.25 &36.5 &17.8 &0.42 &1.91 & 42.3 &193.0\\
  14 & 27. & 14. &-87.50 &35.9 &18.6 &0.28 &1.07 & 35.5 & 40.4\\
  15 &  9. &-30. &-88.00 &35.4 &15.4 &0.32 &1.34 & 22.8 & 72.4\\
  16 & 30. & 21. &-88.00 &33.6 &16.4 &0.31 &1.20 & 25.8 & 56.2\\
  17 & 12. &-45. &-88.00 &33.5 &15.8 &0.31 &1.53 & 30.8 & 91.4\\
  18 & 39. &  9. &-88.50 &33.3 &17.1 &0.23 &0.72 & 15.2 & 15.0\\
  19 & 17. &  3. &-86.50 &33.0 &17.2 &0.29 &0.98 & 24.8 & 35.1\\
  20 & 30. & 11. &-86.00 &32.3 &17.6 &0.29 &1.27 & 27.3 & 58.9\\
  21 & 11. &  9. &-87.00 &32.0 &14.8 &0.26 &1.64 & 26.1 & 88.1\\
  22 & 23. &  2. &-87.00 &31.7 &14.9 &0.35 &1.25 & 29.8 & 68.9\\
  23 &  8. &-38. &-88.25 &31.6 &13.4 &0.36 &1.34 & 20.3 & 81.4\\
  24 & -8. &  6. &-88.50 &31.6 &17.0 &0.30 &1.35 & 34.0 & 68.9\\
  25 &  6. & 41. &-87.25 &31.2 &14.6 &0.29 &0.75 & 16.3 & 20.6\\
  26 &  3. &  9. &-87.25 &31.0 &17.4 &0.28 &1.36 & 31.5 & 65.3\\
  27 &  9. & 23. &-87.25 &30.7 &15.8 &0.31 &1.22 & 20.3 & 58.1\\
  28 & 14. &-44. &-88.75 &30.5 &15.4 &0.21 &0.78 &  9.7 & 16.1\\
  29 & 44. &-17. &-86.50 &30.2 &18.5 &0.28 &0.80 & 20.1 & 22.6\\
  30 & -3. &  3. &-87.75 &30.2 &17.2 &0.32 &1.52 & 31.7 & 93.2\\
  31 & 12. &-27. &-87.50 &30.0 &17.1 &0.42 &0.90 & 35.5 & 42.9\\
  32 & 39. &-18. &-84.75 &29.5 &15.7 &0.26 &0.93 & 20.0 & 28.3\\
  33 & 11. & 23. &-88.00 &28.8 &16.1 &0.32 &1.20 & 29.2 & 58.1\\
  34 &  5. & 44. &-87.00 &28.2 &16.2 &0.27 &0.94 & 14.5 & 30.1\\
  35 & 35. &-12. &-85.75 &28.1 &13.4 &0.28 &1.28 & 18.4 & 57.8\\
  36 & 38. &-24. &-85.75 &28.0 &13.4 &0.19 &0.95 & 10.3 & 21.6\\
  37 & 26. & 35. &-86.50 &27.8 &16.1 &0.43 &1.51 & 59.1 &124.0\\
  38 & 45. &-41. &-86.00 &27.5 &17.3 &0.20 &0.86 & 10.1 & 18.6\\
  39 &  6. & 18. &-88.00 &27.1 &15.7 &0.26 &0.98 & 20.8 & 31.5\\
  40 & 38. & 12. &-85.50 &27.0 &14.6 &0.25 &1.12 & 17.7 & 39.5\\
  41 & 42. & 15. &-85.50 &26.9 &16.6 &0.27 &1.41 & 27.4 & 67.6\\
  42 & 39. &-26. &-89.25 &26.8 &12.9 &0.37 &1.12 & 20.2 & 58.5\\
  43 & 39. &-17. &-85.50 &26.8 &14.9 &0.24 &1.00 & 13.7 & 30.2\\
  44 & 38. & 17. &-88.25 &26.8 &15.9 &0.26 &0.76 & 15.8 & 18.9\\
  45 & 47. & 11. &-88.75 &26.7 &14.4 &0.18 &1.36 & 15.9 & 41.9\\
  46 & 30. & -3. &-86.25 &26.7 &16.0 &0.30 &1.37 & 31.4 & 70.9\\
  47 & 35. &  9. &-84.75 &26.4 &15.5 &0.37 &0.96 & 28.8 & 43.0\\
  48 & 44. &-38. &-85.50 &26.4 &16.1 &0.22 &0.92 & 14.3 & 23.5\\
  49 & 20. & 20. &-87.75 &26.3 &15.3 &0.31 &1.47 & 34.8 & 84.4\\
  50 & 30. &-45. &-84.00 &26.0 &12.8 &0.30 &1.22 & 13.9 & 56.3\\
\noalign{\smallskip}
\hline
\end{tabular}
\end{flushleft}
\end{table*}

\addtocounter{table}{-1}

\clearpage
\begin{table*}
\caption[]{({\it continued})
}
\begin{flushleft}
\begin{tabular}{rrrcrrcccr}
\hline\noalign{\smallskip}
\multicolumn{1}{c}{Clump}& \multicolumn{1}{c}{$\Delta$RA}& 
\multicolumn{1}{c}{$\Delta$Dec}&
\multicolumn{1}{c}{$V_{\rm lsr}$} & \multicolumn{1}{c}{$T_{\rm peak}$} &
$T_{\rm ave}$ & $r_{\rm eff}$ & $\Delta v_{\rm fwhm}$ & $M_{\rm CO}$ &
$M_{\rm vir}$ \\
\multicolumn{1}{l}{}& \multicolumn{1}{c}{[\arcsec]}&
\multicolumn{1}{c}{[\arcsec]} & [km$\,$s$^{-1}$] & \multicolumn{1}{c}{[K]} &
\multicolumn{1}{c}{[K]} & [pc] & [km$\,$s$^{-1}$] & 
[M$_\odot$] & [M$_\odot$]\\
\multicolumn{1}{c}{(1)} & \multicolumn{1}{c}{(2)} & \multicolumn{1}{c}{(3)} &
\multicolumn{1}{c}{(4)} & \multicolumn{1}{c}{(5)} & \multicolumn{1}{c}{(6)} &
\multicolumn{1}{c}{(7)} & \multicolumn{1}{c}{(8)} & \multicolumn{1}{c}{(9)} &
\multicolumn{1}{c}{(10)} \\
\hline\noalign{\smallskip}
  51 & 14. &  5. &-84.00 &25.7 &15.7 &0.49 &1.08 & 37.1 & 72.0\\
  52 &-14. &-33. &-87.00 &25.3 &12.7 &0.27 &0.59 &  8.4 & 11.8\\
  53 & 42. & 41. &-87.25 &24.9 &12.8 &0.30 &1.02 & 11.7 & 39.3\\
  54 & 30. & -5. &-84.50 &23.7 &12.1 &0.33 &0.88 & 15.8 & 32.2\\
  55 & 44. & 41. &-85.00 &22.8 &12.6 &0.30 &0.96 &  7.9 & 34.8\\
  56 & 44. &-33. &-83.50 &22.7 &12.6 &0.22 &0.53 &  4.2 &  7.8\\
  57 & 44. &  5. &-84.00 &22.5 &12.9 &0.31 &1.34 & 20.2 & 70.1\\
  58 & 39. & 41. &-84.00 &22.1 &13.5 &0.10 &0.69 &  2.7 &  6.0\\
  59 & 36. & 36. &-84.00 &21.7 &12.6 &0.23 &1.40 &  9.8 & 56.8\\
  60 & 15. & 38. &-85.00 &21.5 &12.1 &0.39 &1.48 & 18.2 &108.0\\
  61 & 27. & -8. &-85.25 &21.4 &12.0 &0.17 &1.06 &  5.3 & 24.1\\
  62 & 41. &-26. &-84.75 &21.3 &13.1 &0.22 &1.06 & 10.7 & 31.1\\
  63 & 18. &-38. &-84.75 &21.2 &12.0 &0.31 &1.86 & 14.0 &135.0\\
  64 & 21. &-27. &-85.50 &21.1 &12.3 &0.21 &1.01 &  6.4 & 27.0\\
  65 & 44. & 24. &-84.00 &21.0 &12.5 &0.28 &1.28 & 13.2 & 57.8\\
  66 & 32. &-30. &-85.50 &20.9 &12.4 &0.23 &0.83 &  8.7 & 20.0\\
  67 & 24. & 15. &-85.00 &20.9 &12.3 &0.31 &0.80 & 12.6 & 25.0\\
  68 & 12. &  0. &-82.75 &20.8 &12.7 &0.20 &0.68 &  4.8 & 11.7\\
  69 & 33. &-33. &-83.50 &20.7 &12.9 &0.18 &1.08 &  3.7 & 26.5\\
  70 &  9. &-12. &-88.00 &20.4 &12.7 &0.37 &1.11 & 24.0 & 57.4\\
  71 & 18. &  2. &-85.50 &20.4 &12.4 &0.34 &0.99 & 14.3 & 42.0\\
  72 & 15. & 18. &-86.75 &20.4 &13.0 &0.31 &1.08 & 15.6 & 45.6\\
  73 & 45. & 15. &-83.25 &20.3 &12.5 &0.22 &1.31 &  6.5 & 47.6\\
  74 & -6. & 23. &-87.75 &20.2 &12.6 &0.13 &0.44 &  2.0 &  3.2\\
  75 & 29. & 15. &-88.50 &19.9 &12.3 &0.29 &0.77 &  9.0 & 21.7\\
  76 & 26. & 15. &-85.50 &19.9 &13.5 &0.23 &0.53 &  6.2 &  8.1\\
  77 & 23. &-36. &-88.25 &19.9 &12.3 &0.33 &1.03 & 15.0 & 44.1\\
  78 & 39. &-39. &-89.00 &19.8 &11.9 &0.28 &0.90 &  7.6 & 28.6\\
  79 & 42. & 30. &-83.25 &19.6 &12.2 &0.23 &0.92 &  6.4 & 24.5\\
  80 & 23. &-15. &-86.25 &19.6 &11.7 &0.29 &1.13 & 12.2 & 46.7\\
  81 & 12. &  2. &-83.25 &19.5 &12.9 &0.44 &1.23 & 20.3 & 83.9\\
  82 & 30. &-23. &-86.25 &19.4 &12.6 &0.30 &1.53 & 19.9 & 88.5\\
  83 & 27. &-12. &-89.00 &19.1 &12.0 &0.34 &1.27 & 14.6 & 69.1\\
  84 & 39. &  6. &-89.50 &19.0 &11.9 &0.35 &0.66 &  8.8 & 19.2\\
  85 &  2. & 14. &-89.00 &18.7 &12.2 &0.32 &0.95 & 13.8 & 36.4\\
  86 & 27. & 26. &-84.00 &18.7 &11.9 &0.39 &1.57 & 21.7 &121.0\\
  87 & 12. &-18. &-89.00 &18.6 &12.6 &0.24 &1.06 &  6.8 & 34.0\\
  88 & 21. & -6. &-85.75 &18.6 &11.7 &0.27 &1.37 &  9.4 & 63.9\\
  89 & 14. &-27. &-86.50 &18.5 &11.6 &0.48 &1.30 & 19.0 &102.0\\
  90 & 21. &-42. &-88.50 &18.4 &12.3 &0.18 &1.18 &  6.7 & 31.6\\
  91 & 20. &-11. &-88.75 &18.4 &11.7 &0.28 &0.85 &  9.2 & 25.5\\
  92 &-27. &-30. &-89.75 &18.2 &11.6 &0.29 &0.65 &  5.2 & 15.4\\
  93 & 21. & 11. &-85.25 &18.0 &12.4 &0.29 &1.08 & 17.0 & 42.6\\
  94 & 18. &-15. &-89.25 &17.9 &11.9 &0.23 &0.99 &  9.1 & 28.4\\
  95 & 18. &-11. &-89.25 &17.9 &12.5 &0.24 &0.91 &  7.4 & 25.0\\
  96 & 27. &-12. &-85.00 &17.8 &10.9 &0.20 &0.79 &  5.0 & 15.7\\
  97 & -2. & 30. &-87.50 &17.5 &12.0 &0.32 &1.44 & 10.0 & 83.6\\
\noalign{\smallskip}
\hline
\end{tabular}
\end{flushleft}
\end{table*}

\begin{table*}
\caption[]{CS Clump parameters of WB89~380.
Right ascension and declination are in
arcsecond offsets from the central position of the map: 
$(\alpha,\delta)_{1950}$ = (1$^h$ 4$^m$ 35\fs7 , +65\degr\ 5\arcmin\ 21\arcsec).
\label{380cscl}}
\begin{flushleft}
\begin{tabular}{rrrcrrccr}
\hline\noalign{\smallskip}
\multicolumn{1}{c}{Clump}& \multicolumn{1}{c}{$\Delta$RA}& 
\multicolumn{1}{c}{$\Delta$Dec}&
\multicolumn{1}{c}{$V_{\rm lsr}$} & \multicolumn{1}{c}{$T_{\rm peak}$} &
$T_{\rm ave}$ & $r_{\rm eff}$ & $\Delta v_{\rm fwhm}$ & $M_{\rm vir}$ \\
\multicolumn{1}{l}{}& \multicolumn{1}{c}{[\arcsec]}&
\multicolumn{1}{c}{[\arcsec]} & [km$\,$s$^{-1}$] & \multicolumn{1}{c}{[K]} &
\multicolumn{1}{c}{[K]} & [pc] & [km$\,$s$^{-1}$] & [M$_\odot$]\\
\multicolumn{1}{c}{(1)} & \multicolumn{1}{c}{(2)} & \multicolumn{1}{c}{(3)} &
\multicolumn{1}{c}{(4)} & \multicolumn{1}{c}{(5)} & \multicolumn{1}{c}{(6)} &
\multicolumn{1}{c}{(7)} & \multicolumn{1}{c}{(8)} & \multicolumn{1}{c}{(9)} \\
\hline\noalign{\smallskip}
   1 &  0. &  0. &-86.50 & 6.4 & 3.1 &0.45 &1.57 &139.8\\
   2 & -2. & 12. &-86.80 & 6.4 & 3.1 &0.51 &1.41 &127.8\\
   3 &  2. & 10. &-86.20 & 6.3 & 3.5 &0.38 &1.31 & 82.2\\
   4 &  4. &  6. &-85.30 & 5.8 & 3.1 &0.50 &1.09 & 74.9\\
   5 &  4. & -4. &-88.00 & 5.4 & 2.9 &0.52 &1.06 & 73.6\\
   6 & 10. &-14. &-87.10 & 5.1 & 2.4 &0.54 &1.45 &143.1\\
   7 &  8. & 24. &-88.00 & 3.9 & 2.2 &0.46 &1.16 & 78.0\\
   8 & 10. &  0. &-83.80 & 3.8 & 2.1 &0.49 &1.81 &202.3\\
   9 & 22. & 16. &-85.60 & 3.4 & 2.1 &0.42 &2.42 &309.9\\
  10 &-16. &-14. &-87.40 & 3.4 & 2.0 &0.49 &2.28 &320.9\\
  11 &-18. &  6. &-87.10 & 3.4 & 2.2 &0.36 &1.51 &103.4\\
  12 &  4. & 24. &-85.90 & 3.3 & 2.0 &0.42 &1.90 &191.0\\
  13 & 20. &-12. &-86.80 & 3.2 & 2.1 &0.36 &2.74 &340.5\\
  14 &-16. &  8. &-91.60 & 3.1 & 2.1 &0.19 &0.39 &  3.6\\
  15 & 10. &-14. &-86.50 & 3.1 & 2.1 &0.37 &2.06 &197.8\\
  16 &-16. & 22. &-88.30 & 3.1 & 2.2 &0.42 &1.47 &114.4\\
\noalign{\smallskip}
\hline
\end{tabular}
\end{flushleft}
\end{table*}

\begin{table*}
\caption[]{CO and CS Clump parameters of WB89~437.
Right ascension and declination are in
arcsecond offsets from the central position of the map: 
$(\alpha,\delta)_{1950}$ = (2$^h$ 39$^m$ 30\fs6 , 
+62\degr\ 44\arcmin\ 22\arcsec).
\label{437cocscl}}
\begin{flushleft}
\begin{tabular}{rrrcrrcccr}
\hline\noalign{\smallskip}
\multicolumn{1}{c}{Clump}& \multicolumn{1}{c}{$\Delta$RA}& 
\multicolumn{1}{c}{$\Delta$Dec}&
\multicolumn{1}{c}{$V_{\rm lsr}$} & \multicolumn{1}{c}{$T_{\rm peak}$} &
$T_{\rm ave}$ & $r_{\rm eff}$ & $\Delta v_{\rm fwhm}$ & $M_{\rm CO}$ &
$M_{\rm vir}$ \\
\multicolumn{1}{l}{}& \multicolumn{1}{c}{[\arcsec]}&
\multicolumn{1}{c}{[\arcsec]} & [km$\,$s$^{-1}$] & \multicolumn{1}{c}{[K]} &
\multicolumn{1}{c}{[K]} & [pc] & [km$\,$s$^{-1}$] & 
[M$_\odot$] & [M$_\odot$]\\
\multicolumn{1}{c}{(1)} & \multicolumn{1}{c}{(2)} & \multicolumn{1}{c}{(3)} &
\multicolumn{1}{c}{(4)} & \multicolumn{1}{c}{(5)} & \multicolumn{1}{c}{(6)} &
\multicolumn{1}{c}{(7)} & \multicolumn{1}{c}{(8)} & \multicolumn{1}{c}{(9)} &
\multicolumn{1}{c}{(10)} \\
\hline\noalign{\smallskip}
CO & & & & & & & & & \\
   2 & 12. & 15. &-72.00 &33.0 &12.1 &0.85 &1.30 &308.0 & 181.0\\
   3 &  3. &  6. &-73.00 &28.5 &13.3 &0.78 &1.55 &327.0 & 236.0\\
   4 & 24. & 15. &-73.25 &25.3 & 8.6 &0.86 &1.57 &267.0 & 267.0\\
   5 &  0. &-27. &-72.50 &21.2 & 7.4 &0.70 &1.56 &206.0 & 215.0\\
   6 &-27. & 18. &-69.75 &18.4 & 8.5 &1.10 &2.95 &480.0 &1206.2\\
   7 &-24. &-12. &-72.25 &18.3 & 8.3 &0.73 &1.80 &222.0 & 298.0\\
   8 &  0. &  6. &-75.75 & 9.0 & 3.5 &0.83 &1.80 & 63.4 & 339.0\\
   9 & -6. &  9. &-65.25 & 6.1 & 2.9 &0.90 &2.22 & 60.8 & 559.0\\
  10 &  0. &  6. &-78.25 & 5.6 & 2.6 &0.71 &1.19 & 24.0 & 127.0\\
  11 &  6. &-15. &-75.50 & 4.6 & 2.5 &0.45 &1.55 & 13.6 & 136.0\\
  12 &  9. &  9. &-80.25 & 3.9 & 2.2 &0.60 &1.01 &  9.6 &  77.1\\
CS & & & & & & & & & \\
   1 &  3. &  6. &-71.60 & 6.0 & 1.8 &1.11 &1.71 & -- &409.0\\
   2 &  6. &  6. &-72.50 & 4.2 & 1.4 &1.02 &1.23 & -- &194.4\\
\noalign{\smallskip}
\hline
\end{tabular}
\end{flushleft}
\end{table*}

\begin{table*}
\caption[]{Clump parameters of RMC low-resolution $^{13}$CO data.
Gal. longitude and latitude are in
arcmin offsets from the central position of the map: 
{\sl l, b} = 207\pad 5, $-$1\pad 823
\label{rmclowcocl}}
\begin{flushleft}
\begin{tabular}{rrrcrrcccr}
\hline\noalign{\smallskip}
\multicolumn{1}{c}{Clump}& \multicolumn{1}{c}{$\Delta l$}& 
\multicolumn{1}{c}{$\Delta b$}&
\multicolumn{1}{c}{$V_{\rm lsr}$} & \multicolumn{1}{c}{$T_{\rm peak}$} &
$T_{\rm ave}$ & $r_{\rm eff}$ & $\Delta v_{\rm fwhm}$ & $M_{\rm lte}$ &
$M_{\rm vir}$ \\
\multicolumn{1}{l}{}& \multicolumn{1}{c}{[\arcmin]}&
\multicolumn{1}{c}{[\arcmin]} & [km$\,$s$^{-1}$] & \multicolumn{1}{c}{[K]} &
\multicolumn{1}{c}{[K]} & [pc] & [km$\,$s$^{-1}$] & 
[M$_\odot$] & [M$_\odot$]\\
\multicolumn{1}{c}{(1)} & \multicolumn{1}{c}{(2)} & \multicolumn{1}{c}{(3)} &
\multicolumn{1}{c}{(4)} & \multicolumn{1}{c}{(5)} & \multicolumn{1}{c}{(6)} &
\multicolumn{1}{c}{(7)} & \multicolumn{1}{c}{(8)} & \multicolumn{1}{c}{(9)} &
\multicolumn{1}{c}{(10)} \\
\hline\noalign{\smallskip}
   1 &-30. &  0. & 15.60 & 8.6 & 2.1 &2.32 &2.19 &2510.0 &1400\\
   2 &-24. & -2. & 10.84 & 6.3 & 2.0 &2.82 &1.50 &2840.0 & 799\\
   3 &  3. &  6. & 12.88 & 6.1 & 2.0 &3.00 &2.14 &2810.0 &1730\\
   4 &-15. &  0. & 12.88 & 6.0 & 2.2 &2.56 &2.45 &1430.0 &1940\\
   5 &-14. &-20. & 15.60 & 5.9 & 1.6 &2.79 &1.49 &1310.0 & 780\\
   6 &-23. & -5. & 12.88 & 5.5 & 2.4 &1.73 &1.93 &1240.0 & 812\\
   7 & -6. & -8. & 16.28 & 5.3 & 1.4 &2.20 &1.73 & 551.0 & 830\\
   8 &-41. &-11. & 16.28 & 5.2 & 2.0 &1.94 &2.02 &1000.0 & 997\\
   9 &-44. & -8. & 15.60 & 5.1 & 2.0 &1.27 &1.61 & 387.0 & 415\\
  10 &-39. &-33. & 14.24 & 5.1 & 1.7 &1.98 &2.07 & 901.0 &1070\\
  11 &-26. & -3. & 12.88 & 5.0 & 2.1 &2.24 &1.99 &1220.0 &1120\\
  12 &-24. & -3. & 16.28 & 4.5 & 1.4 &1.91 &1.72 & 508.0 & 712\\
  13 &-21. &  2. & 12.20 & 4.5 & 1.8 &1.74 &1.98 & 739.0 & 860\\
  14 &-38. & -5. & 14.92 & 4.5 & 1.6 &2.18 &2.43 &1160.0 &1620\\
  15 & 17. &  3. & 11.52 & 4.3 & 1.9 &2.99 &1.57 &2510.0 & 929\\
  16 & -9. & -5. & 12.20 & 4.1 & 1.6 &2.88 &2.05 &1560.0 &1530\\
  17 & 12. & -6. & 14.92 & 3.9 & 1.6 &2.94 &2.17 &1860.0 &1740\\
  18 &-35. & 14. & 14.24 & 3.2 & 1.4 &0.87 &1.57 & 126.0 & 270\\
  19 &  6. & -5. & 15.60 & 3.1 & 1.2 &2.58 &2.19 & 718.0 &1560\\
  20 & 24. &  2. & 10.84 & 3.1 & 1.2 &2.49 &1.75 & 506.0 & 961\\
  21 & -9. & 24. & 12.20 & 3.0 & 1.2 &2.64 &2.05 & 793.0 &1400\\
  22 &  9. & 15. & 14.92 & 3.0 & 1.1 &2.49 &2.05 & 441.0 &1320\\
  23 &  6. & -8. & 11.52 & 2.9 & 1.1 &2.23 &2.07 & 508.0 &1200\\
  24 &-35. & 11. & 14.92 & 2.8 & 1.0 &1.18 &1.77 & 141.0 & 466\\
  25 &-42. &-42. & 13.56 & 2.8 & 1.1 &1.93 &1.39 & 302.0 & 470\\
  26 &-24. & 11. &  9.48 & 2.7 & 0.9 &1.25 &1.24 & 105.0 & 242\\
  27 & 11. & -6. & 11.52 & 2.5 & 1.3 &2.68 &1.84 & 789.0 &1140\\
  28 & -3. & 33. & 13.56 & 2.3 & 0.8 &2.20 &2.07 & 202.0 &1190\\
  29 & -8. & 33. & 14.24 & 2.3 & 0.9 &1.64 &1.50 & 141.0 & 465\\
  30 &  0. &-14. & 10.16 & 2.2 & 1.0 &2.79 &2.84 & 676.0 &2840\\
  31 &-41. &-15. & 14.24 & 2.1 & 0.9 &2.06 &2.79 & 241.0 &2020\\
  32 & 18. &  3. & 14.92 & 2.0 & 1.0 &2.68 &1.72 & 493.0 & 999\\
  33 & 21. & 14. & 10.84 & 1.9 & 0.8 &1.99 &1.96 & 193.0 & 963\\
  34 & -2. &  6. &  5.40 & 1.9 & 0.9 &1.39 &1.78 & 100.0 & 555\\
  35 &  0. &  3. &  5.40 & 1.6 & 0.9 &1.02 &0.85 &  40.8 &  92.9\\
  36 &-32. &-41. & 11.52 & 1.6 & 0.9 &1.66 &1.41 & 153.0 & 416\\
  37 &-12. & -6. & 16.28 & 1.5 & 0.7 &1.88 &2.54 & 129.0 &1530\\
  38 &  0. & -3. & 10.84 & 1.5 & 0.9 &2.07 &2.41 & 264.0 &1510\\
  39 & 44. &  8. & 11.52 & 1.5 & 0.8 &0.52 &1.48 &  18.9 & 144\\
  40 &-35. &-36. & 14.24 & 1.5 & 0.9 &0.94 &1.48 &  73.9 & 259\\
  41 & -5. & 27. &  6.76 & 1.4 & 0.8 &2.00 &2.13 & 126.0 &1140\\
  42 &-11. &-14. &  8.12 & 1.4 & 0.8 &2.12 &1.99 & 199.0 &1060\\
  43 &-33. &-29. & 16.96 & 1.3 & 0.9 &1.11 &0.85 &  42.4 & 101\\
  44 & 41. & -9. & 10.16 & 1.3 & 0.7 &1.57 &0.89 &  60.9 & 157\\
  45 &-12. &-17. & 12.88 & 1.3 & 0.8 &0.78 &1.14 &  50.2 & 128\\
  46 &-26. &-36. & 14.92 & 1.1 & 0.8 &0.70 &0.79 &  15.0 &  55.1\\
\noalign{\smallskip}
\hline
\end{tabular}
\end{flushleft}
\end{table*}

\begin{table*}
\caption[]{Clump parameters of RMC high-resolution $^{13}$CO data.
Gal. longitude and latitude are in
arcsec offsets from the central position:
{\sl l, b} = 207\pad 0154, $-$1\pad 8228
\label{rmchighcocl}}
\begin{flushleft}
\begin{tabular}{rrrcrrcccr}
\hline\noalign{\smallskip}
\multicolumn{1}{c}{Clump}& \multicolumn{1}{c}{$\Delta l$}& 
\multicolumn{1}{c}{$\Delta b$}&
\multicolumn{1}{c}{$V_{\rm lsr}$} & \multicolumn{1}{c}{$T_{\rm peak}$} &
$T_{\rm ave}$ & $r_{\rm eff}$ & $\Delta v_{\rm fwhm}$ & $M_{\rm lte}$ &
$M_{\rm vir}$ \\
\multicolumn{1}{l}{}& \multicolumn{1}{c}{[\arcsec]}&
\multicolumn{1}{c}{[\arcsec]} & [km$\,$s$^{-1}$] & \multicolumn{1}{c}{[K]} &
\multicolumn{1}{c}{[K]} & [pc] & [km$\,$s$^{-1}$] & 
[M$_\odot$] & [M$_\odot$]\\
\multicolumn{1}{c}{(1)} & \multicolumn{1}{c}{(2)} & \multicolumn{1}{c}{(3)} &
\multicolumn{1}{c}{(4)} & \multicolumn{1}{c}{(5)} & \multicolumn{1}{c}{(6)} &
\multicolumn{1}{c}{(7)} & \multicolumn{1}{c}{(8)} & \multicolumn{1}{c}{(9)} &
\multicolumn{1}{c}{(10)} \\
\hline\noalign{\smallskip}
   1 & -855. & -420. & 15.50 &17.0 & 6.7 &0.57 &1.61 & 123.0 & 186\\
   2 &-1050. & -420. & 20.12 &16.2 & 4.9 &0.34 &1.25 &  35.2 &  66.9\\
   3 & -945. & -300. & 12.42 &11.7 & 5.3 &0.43 &0.96 &  30.1 &  49.9\\
   4 & -945. & -315. & 13.08 &11.7 & 5.6 &0.44 &0.81 &  23.4 &  36.4\\
   5 & -885. & -315. & 13.74 &10.8 & 5.5 &0.65 &0.98 &  59.1 &  78.7\\
   6 & -930. & -345. & 22.10 & 7.8 & 3.2 &0.26 &1.29 &   6.6 &  54.5\\
   7 & -915. & -300. & 21.66 & 7.8 & 3.3 &0.31 &0.98 &   8.0 &  37.5\\
   8 & -810. & -330. & 20.12 & 6.7 & 3.2 &0.22 &0.87 &   3.7 &  21.0\\
   9 & -915. & -390. & 12.64 & 5.8 & 2.6 &0.50 &1.30 &  11.0 & 106\\
  10 &    0. &  -45. & 16.38 &19.9 & 5.8 &0.66 &1.02 &  87.4 &  86.5\\
  11 &   60. &  -75. & 15.06 &19.3 & 6.6 &0.50 &1.56 & 115.0 & 153\\
  12 &   30. &  -30. & 15.06 &17.9 &10.3 &0.41 &1.27 &  88.4 &  83.3\\
  13 &  150. &  -90. & 16.16 &13.5 & 4.3 &0.48 &0.85 &  25.7 &  43.7\\
  14 &   90. &   15. & 15.06 &13.1 & 5.8 &0.47 &1.28 &  48.5 &  97.0\\
  15 &  195. & -105. & 15.94 & 8.1 & 3.1 &0.47 &0.95 &  15.3 &  53.4\\
  16 &  195. & -105. & 11.76 & 8.0 & 3.7 &0.52 &0.94 &  24.4 &  57.9\\
  17 &  105. &   45. & 15.28 & 7.8 & 3.2 &0.41 &1.16 &  14.5 &  69.5\\
  18 &  300. &  -15. & 10.22 & 6.8 & 3.5 &0.50 &1.30 &  19.2 & 106.0\\
  19 &  210. &   15. & 11.98 & 6.7 & 3.2 &0.35 &0.92 &   6.9 &  37.3\\
  20 &  315. &  -75. & 10.88 & 6.7 & 3.3 &0.44 &1.16 &  13.0 &  74.6\\
  21 &  165. &    0. & 11.76 & 6.6 & 3.1 &0.45 &0.99 &  11.4 &  55.6\\
  22 &   30. &  -75. & 10.66 & 6.6 & 3.3 &0.44 &0.97 &  12.9 &  52.2\\
  23 &  240. &  -30. & 10.66 & 6.3 & 3.3 &0.52 &0.82 &  17.1 &  44.1\\
  24 &  240. & -105. & 12.86 & 6.2 & 3.3 &0.60 &1.01 &  30.6 &  77.1\\
  25 &  105. &  -75. & 10.22 & 6.0 & 3.1 &0.34 &0.93 &   7.2 &  37.1\\
  26 &  225. &  -30. & 11.76 & 5.6 & 2.9 &0.37 &0.89 &   9.2 &  36.9\\
  27 &  150. &  -90. & 11.10 & 5.5 & 3.1 &0.50 &0.81 &  13.8 &  41.3\\
  28 &  120. &    0. & 10.00 & 5.5 & 2.9 &0.43 &0.94 &  10.5 &  47.9\\
  29 &   75. &  -15. & 10.00 & 5.4 & 2.6 &0.31 &1.00 &   4.4 &  39.1\\
  30 &  135. &  -45. & 10.66 & 5.3 & 3.4 &0.44 &0.81 &  10.6 &  36.4\\
  31 &   75. & -105. & 10.44 & 5.2 & 3.4 &0.34 &1.07 &   9.5 &  49.0\\
  32 &  210. &    0. & 13.96 & 4.5 & 2.1 &0.41 &1.20 &   4.3 &  74.4\\
  33 &  105. &   45. &  9.78 & 4.3 & 2.3 &0.35 &1.07 &   5.3 &  50.5\\
  34 &  225. &  -30. & 14.18 & 4.2 & 2.1 &0.42 &1.25 &   5.7 &  82.7\\
  35 &  240. &  -90. &  9.34 & 3.2 & 1.9 &0.34 &1.29 &   3.2 &  71.3\\
  36 &   30. &   15. & 10.22 & 3.2 & 1.8 &0.37 &0.92 &   3.5 &  39.5\\
  37 &-1440. &    0. & 10.00 &12.7 & 5.7 &0.44 &1.44 &  49.9 & 115\\
  38 &-1455. &  -45. & 10.00 &11.8 & 4.6 &0.41 &1.75 &  44.4 & 158\\
  39 &-1335. &   45. &  9.56 & 8.3 & 3.7 &0.55 &1.15 &  25.2 &  91.6\\
\noalign{\smallskip}
\hline
\end{tabular}
\end{flushleft}
\end{table*}

\begin{table*}
\caption[]{Clump Parameters of the Orion B South $^{13}$CO data.
Right ascension and declination are in
arcmin offsets from the central position of the map: 
$(\alpha,\delta)_{1950}$ = (5$^h$ 39$^m$ 12\fs0 , $-$1\degr\ 58\arcmin\ 
00\arcsec).
\label{oricocl}}
\begin{flushleft}
\begin{tabular}{rrrcrrcccr}
\hline\noalign{\smallskip}
\multicolumn{1}{c}{Clump}& \multicolumn{1}{c}{$\Delta$RA}& 
\multicolumn{1}{c}{$\Delta$Dec}&
\multicolumn{1}{c}{$V_{\rm lsr}$} & \multicolumn{1}{c}{$T_{\rm peak}$} &
$T_{\rm ave}$ & $r_{\rm eff}$ & $\Delta v_{\rm fwhm}$ & $M_{\rm lte}$ &
$M_{\rm vir}$ \\
\multicolumn{1}{l}{}& \multicolumn{1}{c}{[\arcmin]}&
\multicolumn{1}{c}{[\arcmin]} & [km$\,$s$^{-1}$] & \multicolumn{1}{c}{[K]} &
\multicolumn{1}{c}{[K]} & [pc] & [km$\,$s$^{-1}$] & 
[M$_\odot$] & [M$_\odot$]\\
\multicolumn{1}{c}{(1)} & \multicolumn{1}{c}{(2)} & \multicolumn{1}{c}{(3)} &
\multicolumn{1}{c}{(4)} & \multicolumn{1}{c}{(5)} & \multicolumn{1}{c}{(6)} &
\multicolumn{1}{c}{(7)} & \multicolumn{1}{c}{(8)} & \multicolumn{1}{c}{(9)} &
\multicolumn{1}{c}{(10)} \\
\hline\noalign{\smallskip}
   1 &  1. &  2. & 10.68 &15.5 & 3.6 &1.57 &2.88 &891.0 &1641\\
   2 & -1. &-21. & 10.00 &13.5 & 3.9 &1.56 &2.40 &828.0 &1132\\
   3 & -7. & 10. &  8.87 &10.7 & 4.8 &1.10 &1.90 &314.0 &500\\
   4 &  3. & 18. &  9.10 & 9.3 & 4.8 &0.81 &1.21 &101.0 &149\\
   5 &  1. & 14. & 10.00 & 7.8 & 3.1 &1.06 &2.59 &215.0 &896\\
   6 & 10. & -2. &  9.78 & 6.7 & 3.0 &1.29 &2.33 &304.0 &882\\
   7 &  1. & 25. &  8.65 & 6.3 & 2.8 &0.75 &1.42 & 62.6 &191\\
   8 &  9. & -8. & 10.45 & 6.2 & 3.3 &0.70 &1.40 & 59.6 &173\\
   9 &-11. &-13. & 11.58 & 5.4 & 2.0 &1.36 &2.22 &146.0 &845\\
  10 &  9. &  7. &  9.55 & 5.3 & 3.3 &0.71 &1.33 & 31.0 &158\\
  11 &  9. & 23. &  9.55 & 5.0 & 2.3 &0.81 &1.64 & 62.7 &275\\
  12 &  1. & 25. & 10.91 & 5.0 & 2.4 &1.10 &2.34 &112.0 &759\\
  13 & -6. & 21. &  9.10 & 4.6 & 2.3 &0.85 &1.62 & 54.4 &281\\
  14 &  9. &-31. & 10.23 & 3.7 & 2.3 &1.28 &2.01 & 90.2 &652\\
  15 &  9. &-28. & 10.00 & 3.7 & 2.0 &0.70 &1.66 & 25.9 &243\\
  16 & 11. &-26. &  9.32 & 3.3 & 1.7 &1.08 &1.87 & 64.2 &476\\
  17 & 18. & 20. & 10.45 & 3.2 & 1.3 &1.19 &2.41 & 54.9 &871\\
  18 & 14. & 30. &  9.55 & 3.1 & 1.5 &0.82 &1.63 & 37.5 &275\\
  19 & 19. & 22. &  9.78 & 3.0 & 1.5 &0.85 &2.44 & 44.2 &638\\
  20 & 18. &  8. &  9.55 & 2.6 & 1.2 &0.74 &2.64 & 28.7 &650\\
  21 & 10. & 26. &  7.97 & 2.6 & 1.1 &0.89 &1.15 & 15.1 &148\\
  22 & 16. &  8. & 10.00 & 2.5 & 1.5 &0.88 &1.09 & 13.6 &132\\
  23 &  9. & 25. &  3.90 & 2.5 & 1.1 &0.40 &0.70 &  2.9 & 24.7\\
  24 &-11. & 19. & 10.23 & 2.4 & 1.4 &0.80 &2.11 & 31.6 &449\\
  25 &  4. & 25. & 13.84 & 2.4 & 1.0 &0.45 &1.20 &  3.7 & 81.6\\
  26 &  8. & -1. & 13.39 & 1.9 & 1.2 &0.60 &2.02 &  5.9 &308\\
  27 &  8. &  7. &  7.06 & 1.8 & 1.1 &0.69 &1.08 &  7.0 &101\\
  28 & 19. & 22. &  3.90 & 1.8 & 0.9 &0.58 &0.94 &  3.3 & 64.6\\
  29 &  9. & -3. & 12.26 & 1.8 & 1.0 &0.72 &1.57 &  7.6 &224\\
  30 & 13. & 17. &  8.65 & 1.8 & 1.3 &0.56 &1.31 &  4.0 &121\\
  31 &  7. &  9. &  5.48 & 1.6 & 1.0 &0.51 &0.90 &  3.9 & 52.1\\
  32 & -1. &  7. &  5.48 & 1.5 & 0.8 &0.65 &1.42 &  5.0 &165\\
  33 & 10. & 10. &  5.71 & 1.4 & 0.8 &0.49 &1.11 &  2.5 & 76.1\\
\noalign{\smallskip}
\hline
\end{tabular}
\end{flushleft}
\end{table*}

\end{document}